\documentclass[12pt]{article}
\usepackage{amssymb, latexsym, amsmath, hyperref}

\begin{document}

\title{\textbf{Pseudoduality Between Symmetric Space Sigma Models}}

\author{\textbf{Mustafa Sarisaman}\footnote{msarisaman@physics.miami.edu}}

\date{}

\maketitle
\par
\begin{center}
\textit{Department of Physics\\ University of Miami\\ P.O. Box 248046\\
 Coral Gables, FL 33124 USA}
\end{center}

\

\begin{center}
Monday, April 27, 2009
\end{center}

\

\begin{abstract}
We study the pseudoduality transformation on the symmetric space
sigma models. We switch the Lie group valued pseudoduality equations
to Lie algebra valued ones, which leads to an infinite number of
pseudoduality equations. We obtain an infinite number of conserved
currents on the tangent bundle of the pseudodual manifold. We show
that there can be mixing of decomposed spaces with each other, which
leads to mixings of the following expressions.  We obtain the mixing
forms of curvature relations and one loop renormalization group beta
functions by means of these currents.
\end{abstract}

\vfill

\section{Introduction}\label{sec:int}

We know that there is a well defined duality
transformation\footnote{This transformation is known as
pseudoduality transformation \cite{curtright1, ivanov, alvarez1,
alvarez2}} between target spaces of the sigma models on symmetric
spaces with opposite curvatures which preserves the stress energy
tensors associated with each sigma models though it is not a
canonical transformation. In this paper we present the general
solution of the pseudoduality equations \cite{msarisaman1} between
two symmetric space sigma models, and construct the pseudodual
currents by means of these equations. We will do our calculations
regarding $G$ as a symmetric space $G \times G / G$, and then extend
our construction using Cartan's decomposition of symmetric spaces.
We will use the references \cite{helgason, oneill, wolf,
arvanitoyeorgos} for the symmetric space construction, and utilize
the literature \cite{forger1, forger2, forger3, forger4, evans1,
fendley1, schwarz1, schwarz2, alvaredo1, pousa1, bakas1, evans2,
evans3, evans4} on various applications to sigma models. Since
pseudoduality is defined on spacetime coordinates \cite{alvarez2},
and is best done on the orthonormal coframes bundle \footnote{$SO(M)
= M \times SO(n)$, where $dim(M) = n$.} $SO(M)$, we leave this
construction to later \cite{msarisaman2}. In this paper we will do
our calculations on the pullback bundle of target space $M$. Hence
pulling structures back to spacetime is implicit, and not
emphasized. We will see that this construction will give us
complicated expressions for $T$ as opposed to the simplified form
(identity) on $SO(M)$ \cite{msarisaman2}.

\section{Pseudoduality Between strict WZW Models}
\label{sec:PBSMGC}

We consider a strict WZW sigma model \cite{witten1} based on a
compact Lie group of dimension n. Lagrangian \cite{witten1, evans1,
evans2, evans3, evans4} for this model is defined by
\begin{equation} \label{equation3.1}
 \mathcal{L} =
\frac{1}{2}Tr(g^{-1}\partial_{\mu}gg^{-1}\partial^{\mu}g) + \Gamma
\end{equation}
where $\Gamma$ represents the WZ term, and the field $g$ is given by
the map $g : \Sigma \rightarrow G$. We take $\Sigma$ to be two
dimensional Minkowski space, and $\sigma^{\pm} = \tau \pm \sigma$ is
the standard lightcone coordinates as above. There is a global
continuous symmetry $G_{L}\times G_{R}$ which gives us the conserved
currents $J_{+}^{(L)} = g_{L}^{-1}\partial_{+}g_{L}$ and
$J_{-}^{(R)} = (\partial_{-}g_{R})g_{R}^{-1}$ taking values in the
Lie algebra of $G$, and $g = g_{R}(\sigma^{-})g_{L}(\sigma^{+})$ is
the solution giving the invariance of these currents. The equations
of motion following from (\ref{equation3.1}) correspond to the
conservation of these currents:
\begin{equation} \label{equation3.2}
\partial_{-}(g_{L}^{-1}\partial_{+}g_{L}) =
\partial_{+}[(\partial_{-}g_{R})g_{R}^{-1}] = 0
\end{equation}

Let $\tilde{G}$ be compact Lie group of the same dimension as $G$,
and $\tilde{g} : \Sigma \rightarrow \tilde{G}$. Equations of motion
are given by
\begin{equation} \label{equation3.3}
\partial_{-}(\tilde{g_{L}}^{-1}\partial_{+}\tilde{g_{L}}) =
\partial_{+}[(\partial_{-}\tilde{g_{R}})\tilde{g_{R}}^{-1}] = 0
\end{equation}

Solutions of equations of motion for both models can be combined in
pseudoduality equations as
\begin{align}
(\tilde{g}^{-1}\partial_{+}\tilde{g})^{i} &=
T_{j}^{i}(g^{-1}\partial_{+}g)^{j} \label{equation3.4} \\
(\tilde{g}^{-1}\partial_{-}\tilde{g})^{i} &= -
T_{j}^{i}(g^{-1}\partial_{-}g)^{j} \label{equation3.5}
\end{align}
where $T$ is an orthogonal matrix connecting target space elements
$g^{-1}dg$ and $\tilde{g}^{-1}d\tilde{g}$.

Taking $\partial_{-}$ of first equation (\ref{equation3.4}) with the
help of equations of motions (\ref{equation3.2}) and
(\ref{equation3.3}) shows that $T$ is a function of $\sigma^{+}$
only. Taking $\partial_{+}$ of second equation (\ref{equation3.5})
gives us the following differential equation
\begin{equation} \label{equation3.6}
[(\partial_{+}T)T^{-1}]_{j}^{i} =
f_{ml}^{k}T_{k}^{i}T_{l}^{j}(g_{L}^{-1}\partial_{+}g_{L})^{m} -
\tilde{f}_{kj}^{i}T_{l}^{k}(g_{L}^{-1}\partial_{+}g_{L})^{l}
\end{equation}
We suggest an exponential solution \footnote{We notice that $X \in
so(n)$, the Lie algebra of $SO(n)$} $T = e^{X}$ , and use the result
\cite{helgason, forger1, forger3}
\begin{equation} \label{equation3.7}
(\partial_{+}T)T^{-1} = -\frac{1 - e^{adX}}{adX}\partial_{+}X =
\sum_{n = 0}^{\infty}\frac{1}{(n + 1)!}[X,...,[X, \partial_{+}X]]
\end{equation}
where $adX : \textbf{g}\rightarrow \textbf{g}$, the adjoint
representation of X, and $adX(Y) = [X, Y]$ $ \forall Y \epsilon \
\textbf{g}$. We let $X \rightarrow \varepsilon X$ and look for a
perturbation solution, and hence the left-hand side of equation
(\ref{equation3.6}) is
\begin{equation} \label{equation3.8}
[(\partial_{+}T)T^{-1}]_{j}^{i} = \varepsilon(\partial_{+}X)_{j}^{i}
+ \frac{\varepsilon^{2}}{2}[X, \partial_{+}X]_{j}^{i} +
\frac{\varepsilon^{3}}{3!}[X, [X, \partial_{+}X]] + ...
\end{equation}
We insert an order parameter $\varepsilon$ to the right-hand side of
(\ref{equation3.6}), and get
\begin{align} \label{equation3.9}
[(\partial_{+}T)T^{-1}]_{j}^{i} = &\varepsilon
f_{ml}^{k}T_{k}^{i}T_{l}^{j}(g_{L}^{-1}\partial_{+}g_{L})^{m} -
\varepsilon
\tilde{f}_{kj}^{i}T_{l}^{k}(g_{L}^{-1}\partial_{+}g_{L})^{l} \\
= &\varepsilon f_{ml}^{k}(1 + \varepsilon X)_{k}^{i} (1 +
\varepsilon X)_{l}^{j}(g_{L}^{-1}\partial_{+}g_{L})^{m} -
\varepsilon \tilde{f}_{kj}^{i}
(1 + \varepsilon X)_{l}^{k}(g_{L}^{-1}\partial_{+}g_{L})^{m} \notag \\
= &\varepsilon f_{mj}^{i}(g_{L}^{-1}\partial_{+}g_{L})^{m} -
\varepsilon \tilde{f}_{kj}^{i} (g_{L}^{-1}\partial_{+}g_{L})^{k} +
\varepsilon^{2} f_{ml}^{i}
X_{l}^{j}(g_{L}^{-1}\partial_{+}g_{L})^{m} \notag\\ &+
\varepsilon^{2} f_{mj}^{k}
X_{k}^{i}(g_{L}^{-1}\partial_{+}g_{L})^{m} - \varepsilon^{2}
\tilde{f}_{kj}^{i}X_{l}^{k}(g_{L}^{-1}\partial_{+}g_{L})^{l} +
\mathcal{O}(\varepsilon^{3}) \notag
\end{align}
Comparing (\ref{equation3.8}) and (\ref{equation3.9}) in the first
order of $\varepsilon$ gives us
 \begin{equation} \label{equation3.10}
 (\partial_{+}X)_{j}^{i} = (f_{kj}^{i} -
 \tilde{f}_ {kj}^{i})(g_{L}^{-1}\partial_{+}g_{L})^{k}
 \end{equation}
This leads to the solution
\begin{equation} \label{equation3.11}
X_{j}^{i} = X(0)_{j}^{i} + (f_{kj}^{i} - \tilde{f}_
{kj}^{i})\int_{0}^{\sigma^{+}} (g_{L}^{-1}\partial_{+}g_{L})^{k}
d\sigma'^{+}
\end{equation}
Hence the matrix $T$ may be written as
\begin{equation} \label{equation3.12}
T_{j}^{i} = \delta_{j}^{i} + X(0)_{j}^{i} + (f_{kj}^{i} - \tilde{f}_
{kj}^{i})\int_{0}^{\sigma^{+}} (g_{L}^{-1}\partial_{+}g_{L})^{k}
d\sigma'^{+}
\end{equation}

We see that if both sigma models based on the same groups, i.e $G =
\tilde{G}$, target space of transformed model will be globally
shifted as determined by the tangent space of unit element of $T$.
We set $X(0)_{j}^{i}$ equal to zero.

Now we plug this in the pseudoduality equations (\ref{equation3.4})
and (\ref{equation3.5}) to find fields
$\tilde{g}^{-1}\partial_{+}\tilde{g}$ and
$\tilde{g}^{-1}\partial_{-}\tilde{g}$ which lead us to construct the
pseudodual currents. We switch from Lie group-valued fields to the
lie algebra-valued fields, and we let \footnote{$Y$ is the lie
algebra of $g$, $Y \in \textbf{g}$.} $g = e^{Y}$ and $\tilde{g} =
e^{\tilde{Y}}$ . Using the result \cite{helgason, forger1, forger3}
\begin{equation} \label{equation3.13}
e^{-X}\partial_{\mu}e^{X} = \frac{1 - e^{-adX}}{adX}\partial_{\mu}X
= \sum_{k = 0}^{\infty} \frac{(-1)^{k}}{(k + 1)!}[X,...,[X,
\partial_{\mu}X]]
\end{equation}
we can write the following
\begin{equation} \label{equation3.14}
g_{L}^{-1}\partial_{+}g_{L} = \partial_{+}Y_{L} -
\frac{1}{2!}[Y_{L},
\partial_{+}Y_{L}] + \frac{1}{3!}[Y_{L}, [Y_{L}, \partial_{+} Y_{L}]] + ...
\end{equation}
\begin{align}
g^{-1}\partial_{-}g = &\partial_{-}Y_{R} - [Y_{L},
\partial_{-}Y_{R}] - \frac{1}{2}[Y_{R},
\partial_{-}Y_{R}] + \frac{1}{2}[Y_{L}, [Y_{R}, \partial_{-} Y_{R}]]\label{equation3.15} \\ &+ \frac{1}{2}[Y_{L}, [Y_{L}, \partial_{-} Y_{R}]] + \frac{1}{6}[Y_{R}, [Y_{R},
\partial_{-}Y_{R}]]... \notag
\end{align}
and the equations of motion for the left and right currents will be
\begin{equation} \label{equation3.16}
\partial_{-}(g_{L}^{-1}\partial_{+}g_{L}) = \partial_{+-}^{2}Y_{L} -
\frac{1}{2!}\partial_{-}[Y_{L}, \partial_{+}Y_{L}] +
\frac{1}{3!}\partial_{-}[Y_{L}, [Y_{L}, \partial_{+}Y_{L}]] + ... =
0
\end{equation}
\begin{equation} \label{equation3.17}
\partial_{+}[(\partial_{-}g_{R})g_{R}^{-1}] = \partial_{+-}^{2}Y_{R}
+ \frac{1}{2!}\partial_{+}[Y_{R}, \partial_{-}Y_{R}] +
\frac{1}{3!}\partial_{+}[Y_{R}, [Y_{R}, \partial_{-}Y_{R}]] + ... =
0
\end{equation}
where $g_{L/R} = e^{Y_{L/R}}$, and we used equation
(\ref{equation3.7}). We may write similar equations with tilde $(\
\tilde{}\ )$. Hence transformation matrix $T$ (\ref{equation3.12})
will be
\begin{equation} \label{equation3.18}
T_{j}^{i} = \delta_{j}^{i} + (f_{kj}^{i} -
\tilde{f}_{kj}^{i})Y_{L}^{k} - \frac{1}{2!}(f_{kj}^{i} -
\tilde{f}_{kj}^{i})\int_{0}^{\sigma^{+}}[Y_{L},
\partial_{+}Y_{L}]^{k}d\sigma'^{+}
\end{equation}

We impose a solution $Y = \sum_{n = 1}^{\infty}
\varepsilon^{n}y_{n}$ to determine the nonlinear parts of the
equations (\ref{equation3.14}) and (\ref{equation3.15}) in terms of
$\varepsilon$, where $\varepsilon$ is a small parameter. Thus
transformation matrix (\ref{equation3.18}) becomes
 \begin{equation} \label{equation3.19}
 T_{j}^{i} = \delta_{j}^{i} + \varepsilon (f_{kj}^{i} -
 \tilde{f}_{kj}^{i})y_{L1}^{k} + \varepsilon^{2} (f_{kj}^{i} -
 \tilde{f}_{kj}^{i})[y_{L2}^{k} - \frac{1}{2}\int_{0}^{\sigma^{+}} [y_{L1}, \partial_{+}y_{L1}]^{k} d\sigma'^{+}
 ] + \mathcal{O}(\varepsilon^{3})
 \end{equation}
and we have the following expressions for (\ref{equation3.14}) and
(\ref{equation3.15})
\begin{align}
g_{L}^{-1}\partial_{+}g_{L} &= \varepsilon \partial_{+}y_{L1} +
\varepsilon^{2} (\partial_{+}y_{L2} - \frac{1}{2}[y_{L1},
\partial_{+}y_{L1}]) \label{equation3.20} \\ &+ \varepsilon^{3}(\partial_{+}y_{L3} - \frac{1}{2}[y_{L1}, \partial_{+}y_{L2}] - \frac{1}{2}[y_{L2}, \partial_{+}y_{L1}] + \frac{1}{6}[y_{L1}, [y_{L1},
\partial_{+}y_{L1}]]) + H.O (\varepsilon)\notag
\end{align}
\begin{align}
g^{-1}\partial_{-}g &= \varepsilon \partial_{-}y_{R1} +
\varepsilon^{2}(\partial_{-}y_{R2} - [y_{L1}, \partial_{-}y_{R1}] -
\frac{1}{2}[y_{R1}, \partial_{-}y_{R1}]) \label{equation3.21} \\ &+
\varepsilon^{3}(\partial_{-}y_{R3} - [y_{L2}, \partial_{-}y_{R1}] -
[y_{L1}, \partial_{-}y_{R2}] - \frac{1}{2}[y_{R2},
\partial_{-}y_{R1}] - \frac{1}{2}[y_{R1}, \partial_{-}y_{R2}] \notag \\ &+ \frac{1}{2}[y_{L1}, [y_{R1}, \partial_{-}y_{R1}]] + \frac{1}{2}[y_{L1}, [y_{L1},
\partial_{-}y_{R1}]]) + H.O (\varepsilon) \notag
\end{align}
Therefore first pseudoduality equation (\ref{equation3.4}) can be
split into infinite number of equations, determined by each order of
$\varepsilon$ as follows,

\begin{align} \label{equation3.22}
&(1.i)\ \partial_{+}\tilde{y}_{L1}^{i} = \partial_{+}y_{L1}^{i} \\
&(1.ii)\ \partial_{+}\tilde{y}_{L2}^{i} +
\frac{1}{2}[\tilde{y}_{L1},
\partial_{+}\tilde{y}_{L1}]_{\tilde{G}}^{i} = \partial_{+}y_{L2}^{i} +
\frac{1}{2}[y_{L1}, \partial_{+}y_{L1}]_{G}^{i} \notag \\
&(1.iii)\ \partial_{+}\tilde{y}_{3}^{i} - \frac{1}{2}[\tilde{y}_{1},
\partial_{+}\tilde{y}_{2}]_{\tilde{G}}^{i} - \frac{1}{2}[\tilde{y}_{2},
\partial_{+}\tilde{y}_{1}]_{\tilde{G}}^{i} + \frac{1}{6}[\tilde{y}_{1}, [\tilde{y}_{1},
\partial_{+}\tilde{y}_{1}]_{\tilde{G}}]_{\tilde{G}}^{i} = \partial_{+}y_{3}^{i} \notag \\
&\ + \frac{1}{2}[y_{1}, \partial_{+}y_{2}]_{G}^{i} +
\frac{1}{2}[y_{2}, \partial_{+}y_{1}]_{G}^{i} - [y_{1},
\partial_{+}y_{2}]_{\tilde{G}}^{i} - [y_{2},
\partial_{+}y_{1}]_{\tilde{G}}^{i} - \frac{1}{3}[y_{1}, [y_{1},
\partial_{+}y_{1}]_{G}]_{G}^{i} \notag \\ &\ + \frac{1}{2}[y_{1}, [y_{1},
\partial_{+}y_{1}]_{G}]_{\tilde{G}}^{i} - \frac{1}{2}[\int_{0}^{\sigma^{+}} [y_{1},
\partial_{+}y_{1}]_{G}\ d\sigma'^{+}, \partial_{+}y_{1}]_{G}^{i} + \frac{1}{2}[\int_{0}^{\sigma^{+}} [y_{1},
\partial_{+}y_{1}]_{G}\ d\sigma'^{+},
\partial_{+}y_{1}]_{\tilde{G}}^{i} \notag \\
& \   \cdot\   \cdot\   \cdot \notag
\end{align}
where we used subindex $G$ ($\tilde{G}$) to represent commutation
relations for the sigma model based on Lie group $G$ ($\tilde{G}$).
(1.i) gives $\tilde{y}_{L1} = y_{L1} + C_{L1}$, where $C_{L1}$ is a
constant, and we set it equal to zero, and leads to (1.ii). Likewise
second pseudoduality equation (\ref{equation3.5}) gives the
following infinite set of equations

\begin{align} \label{equation3.23}
&(2.i)\ \partial_{-}\tilde{y}_{R1}^{i} = - \partial_{-}y_{R1}^{i} \\
&(2.ii)\ \partial_{-}\tilde{y}_{R2}^{i} -
\frac{1}{2}[\tilde{y}_{R1},
\partial_{-}\tilde{y}_{R1}]_{\tilde{G}}^{i} = - \partial_{-}y_{R2}^{i} +
\frac{1}{2}[y_{R1}, \partial_{-}y_{R1}]_{G}^{i} \notag \\
&(2.iii)\   \cdot\   \cdot\   \cdot \notag
\end{align}
where we used (2.i) and (1.i) in (2.ii), and (2.i) leads to
$\tilde{y}_{R1} = - y_{R1} + C_{R1}$, $C_{R1}$ is a constant which
is set to zero. We notice the fact that (\ref{equation3.22}) only
depends on $\sigma^{+}$, and (\ref{equation3.23}) on $\sigma^{-}$
point out pseudodual conserved currents, which can be written as
follows
\begin{equation} \label{equation3.24}
\tilde{J}_{+}^{L}(\sigma^{+}) = \tilde{g}^{-1}\partial_{+}\tilde{g}
= \sum_{n = 1}^{\infty} \varepsilon^{n}
\tilde{J}_{+}^{L[n]}(\sigma^{+})
\end{equation}
\begin{equation} \label{equation3.25}
\tilde{J}_{-}^{R}(\sigma^{-}) =
(\partial_{-}\tilde{g})\tilde{g}^{-1} = \sum_{n = 1}^{\infty}
\varepsilon^{n} \tilde{J}_{-}^{R[n]}(\sigma^{-})
\end{equation}
where each component is determined by the orders of $\varepsilon$'s,
which are given by expression (\ref{equation3.20}) (with tilde). The
nonlocal expressions of currents are determined with the help of
(\ref{equation3.22}) and (\ref{equation3.23})
\begin{equation} \label{equation3.26}
\tilde{J}_{+}^{L[1]}(\sigma^{+}) = \partial_{+}\tilde{y}_{L1}^{i} =
\partial_{+}y_{L1}^{i}
\end{equation}
\begin{equation} \label{equation3.27}
\tilde{J}_{+}^{L[2]}(\sigma^{+}) = \partial_{+}\tilde{y}_{L2} -
\frac{1}{2}[\tilde{y}_{L1},
\partial_{+}\tilde{y}_{L1}]_{\tilde{G}} = \partial_{+}y_{L2}^{i} +
\frac{1}{2}[y_{L1}, \partial_{+}y_{L1}]_{G}^{i} - [y_{L1},
\partial_{+}y_{L1}]_{\tilde{G}}
\end{equation}
\begin{equation}
\   \cdot\   \cdot\   \cdot \notag
\end{equation}

\begin{equation} \label{equation3.28}
\tilde{J}_{-}^{R[1]}(\sigma^{-}) = \partial_{-}\tilde{y}_{R1}^{i} =
- \partial_{-}y_{R1}^{i}
\end{equation}
\begin{equation} \label{equation3.29}
\tilde{J}_{-}^{R[2]}(\sigma^{-}) = \partial_{-}\tilde{y}_{R2} +
\frac{1}{2}[\tilde{y}_{R1},
\partial_{-}\tilde{y}_{R1}]_{\tilde{G}} = - \partial_{-}y_{R2}^{i} +
\frac{1}{2}[y_{R1}, \partial_{-}y_{R1}]_{G}^{i} + [y_{R1},
\partial_{-}y_{R1}]_{\tilde{G}}
\end{equation}
\begin{equation}
\   \cdot\   \cdot\   \cdot \notag
\end{equation}

We see that these currents are conserved,
$\partial_{-}\tilde{J}_{+}^{L} = \partial_{+}\tilde{J}_{-}^{R} = 0$.
It is observed that pseudodual currents are expressed as a nonlocal
function of lie algebra valued fields on $\textbf{g}$. As a result
we obtained a family of nonlocal conserved currents on the WZW model
on $G$. This family is a consequence of infinite set of terms of $T$
which is a function of lie algebra valued fields $\textbf{g}$.

\subsection{An Example}

We consider sigma models based on Lie groups $G = SO(n + 1)$ and
$\tilde{G} = SO(n, 1)$. The corresponding lie algebra are given by

\begin{align} \label{equation3.30}
\textbf{so(n + 1)} = \begin{pmatrix}
              a & b \\
              -b^{t} & c \\
            \end{pmatrix} \ \ \ \ \ \textbf{so(n, 1)} = \begin{pmatrix}
             \tilde{a} & \tilde{b} \\
             \tilde{b}^{t} & \tilde{c} \\
           \end{pmatrix} \ \ \ \ \ \begin{array}{c}
                                     a = \tilde{a} = n \times n \\
                                     b = \tilde{b} = n \times 1 \\
                                     c = \tilde{c} = 1 \times 1
                                   \end{array}
\end{align}

Let $g = e^{Y}$ and $\tilde{g} = e^{\tilde{Y}}$, and fields
$g_{L}^{-1}\partial_{+}g_{L}$ and
$\tilde{g_{L}}^{-1}\partial_{+}\tilde{g_{L}}$ are given by
(\ref{equation3.14}). We get the following expressions
\begin{equation}
Y_{L} = \begin{pmatrix}
              a_{L} & b_{L} \\
              -b_{L}^{t} & c_{L} \\
            \end{pmatrix} \ \ \ \ \ \ \ \ \ \ \ \ \ \ \ \partial_{+}Y_{L} = \begin{pmatrix}
              \partial_{+} a_{L} & \partial_{+} b_{L} \\
              -\partial_{+} b_{L}^{t} & \partial_{+} c_{L} \\
            \end{pmatrix} \notag
\end{equation}
\begin{equation}
\tilde{Y_{L}} = \begin{pmatrix}
             \tilde{a_{L}} & \tilde{b_{L}} \\
             \tilde{b_{L}}^{t} & \tilde{c_{L}} \\
           \end{pmatrix} \ \ \ \ \ \ \ \ \ \ \ \ \ \ \ \partial_{+}\tilde{Y_{L}} = \begin{pmatrix}
             \partial_{+}\tilde{a_{L}} & \partial_{+}\tilde{b_{L}} \\
             \partial_{+}\tilde{b_{L}}^{t} & \partial_{+}\tilde{c_{L}} \\
           \end{pmatrix} \notag
\end{equation}
\begin{equation}
[Y_{L}, \partial_{+}Y_{L}] = \left(\begin{smallmatrix}
                               0 & a_{L}\partial_{+}b_{L} + b_{L}\partial_{+}c_{L} - (\partial_{+}a_{L})b_{L} - (\partial_{+}b_{L})c_{L} \\
                               -b_{L}^{t}(\partial_{+}a_{L}) - c_{L}(\partial_{+}b_{L}^{t}) + (\partial_{+}b_{L}^{t})a_{L} + (\partial_{+}c_{L})b_{L}^{t} & 0 \\
                             \end{smallmatrix} \right) \notag
\end{equation}
\begin{equation}
[\tilde{Y}_{L}, \partial_{+}\tilde{Y}_{L}] =
\left(\begin{smallmatrix}
                               0 & \tilde{a}_{L}\partial_{+}\tilde{b}_{L} + \tilde{b}_{L}\partial_{+}\tilde{c}_{L} - (\partial_{+}\tilde{a}_{L})\tilde{b}_{L} - (\partial_{+}\tilde{b}_{L})\tilde{c}_{L} \\
                               \tilde{b}_{L}^{t}(\partial_{+}\tilde{a}_{L}) + \tilde{c}_{L}(\partial_{+}\tilde{b}_{L}^{t}) - (\partial_{+}\tilde{b}_{L}^{t})\tilde{a}_{L} - (\partial_{+}\tilde{c}_{L})\tilde{b}_{L}^{t} & 0 \\
                             \end{smallmatrix} \right) \notag
\end{equation}

Hence up to the second order terms we get the expressions for the
fields on the target space elements
\begin{equation}
g_{L}^{-1}\partial_{+}g_{L} = \begin{pmatrix}
                                X_{1}& X_{2} \\
                                X_{3}& X_{4} \\
                             \end{pmatrix} + H.O \ \ \ \ \ \ \ \
\tilde{g_{L}}^{-1}\partial_{+}\tilde{g_{L}} = \begin{pmatrix}
                                \tilde{X}_{1}&\tilde{X}_{2}  \\
                                \tilde{X}_{3}& \tilde{X_{4}}  \\
                             \end{pmatrix} + H.O
                             \label{equation3.31}
\end{equation}
where we defined the following
\begin{equation}
X_{1} = \partial_{+}a_{L} \ \ \ \ \ \ \tilde{X}_{1} =
\partial_{+}\tilde{a}_{L} \ \ \ \ X_{4} =
\partial_{+}c_{L} \ \ \ \ \ \tilde{X_{4}} = \partial_{+}\tilde{c}_{L}  \notag
\end{equation}
\begin{equation}
X_{2} = \partial_{+}b_{L}- \frac{a_{L}\partial_{+}b_{L} +
b_{L}\partial_{+}c_{L} - (\partial_{+}a_{L})b_{L} -
(\partial_{+}b_{L})c_{L}}{2} \notag
\end{equation}
\begin{equation}
\tilde{X}_{2} = \partial_{+}\tilde{b}_{L} -
\frac{\tilde{a}_{L}\partial_{+}\tilde{b}_{L} +
\tilde{b}_{L}\partial_{+}\tilde{c}_{L} -
(\partial_{+}\tilde{a}_{L})\tilde{b}_{L} -
(\partial_{+}\tilde{b}_{L})\tilde{c}_{L}}{2} \notag
\end{equation}
\begin{equation}
X_{3} = -\partial_{+}b_{L}^{t} - \frac{-
b_{L}^{t}(\partial_{+}a_{L}) - c_{L}(\partial_{+}b_{L}^{t}) +
(\partial_{+}b_{L}^{t})a_{L} + (\partial_{+}c_{L})b_{L}^{t}}{2}
\notag
\end{equation}
\begin{equation}
\tilde{X}_{3} = \partial_{+}\tilde{b}_{L}^{t} -
\frac{\tilde{b}_{L}^{t}(\partial_{+}\tilde{a}_{L}) +
\tilde{c}_{L}(\partial_{+}\tilde{b}_{L}^{t}) -
(\partial_{+}\tilde{b}_{L}^{t})\tilde{a}_{L} -
(\partial_{+}\tilde{c}_{L})\tilde{b}_{L}^{t}}{2} \notag
\end{equation}

Likewise we get the following expressions related to fields
$g^{-1}\partial_{-}g$ and $\tilde{g}^{-1}\partial_{-}\tilde{g}$
using (\ref{equation3.15})
\begin{equation}
[Y_{L}, \partial_{-}Y_{R}] = \left(\begin{smallmatrix}
                               0 & a_{L}\partial_{-}b_{R} + b_{L}\partial_{-}c_{R} - (\partial_{-}a_{R})b_{L} - (\partial_{-}b_{R})c_{L} \\
                               -b_{L}^{t}(\partial_{-}a_{R}) - c_{L}(\partial_{-}b_{R}^{t}) + (\partial_{-}b_{R}^{t})a_{L} + (\partial_{-}c_{R})b_{L}^{t} & 0 \\
                             \end{smallmatrix} \right) \notag
\end{equation}
\begin{equation}
[Y_{R}, \partial_{-}Y_{R}] = \left(\begin{smallmatrix}
                               0 & a_{R}\partial_{-}b_{R} + b_{R}\partial_{-}c_{R} - (\partial_{-}a_{R})b_{R} - (\partial_{-}b_{R})c_{R} \\
                               -b_{R}^{t}(\partial_{-}a_{R}) - c_{R}(\partial_{-}b_{R}^{t}) + (\partial_{-}b_{R}^{t})a_{R} + (\partial_{-}c_{R})b_{R}^{t} & 0 \\
                             \end{smallmatrix} \right) \notag
\end{equation}
\begin{equation}
[\tilde{Y}_{L}, \partial_{-}\tilde{Y}_{R}] =
\left(\begin{smallmatrix}
                               0 & \tilde{a}_{L}\partial_{-}\tilde{b}_{R} + \tilde{b}_{L}\partial_{-}\tilde{c}_{R} - (\partial_{-}\tilde{a}_{R})\tilde{b}_{L} - (\partial_{-}\tilde{b}_{R})\tilde{c}_{L} \\
                               \tilde{b}_{L}^{t}(\partial_{-}\tilde{a}_{R}) + \tilde{c}_{L}(\partial_{-}\tilde{b}_{R}^{t}) - (\partial_{-}\tilde{b}_{R}^{t})\tilde{a}_{L} - (\partial_{-}\tilde{c}_{R})\tilde{b}_{L}^{t} & 0 \\
                             \end{smallmatrix} \right) \notag
\end{equation}
\begin{equation}
[\tilde{Y}_{R}, \partial_{-}\tilde{Y}_{R}] =
\left(\begin{smallmatrix}
                               0 & \tilde{a}_{R}\partial_{-}\tilde{b}_{R} + \tilde{b}_{R}\partial_{-}\tilde{c}_{R} - (\partial_{-}\tilde{a}_{R})\tilde{b}_{R} - (\partial_{-}\tilde{b}_{R})\tilde{c}_{R} \\
                               \tilde{b}_{R}^{t}(\partial_{-}\tilde{a}_{R}) + \tilde{c}_{R}(\partial_{-}\tilde{b}_{R}^{t}) - (\partial_{-}\tilde{b}_{R}^{t})\tilde{a}_{R} - (\partial_{-}\tilde{c}_{R})\tilde{b}_{R}^{t} & 0 \\
                             \end{smallmatrix} \right) \notag
\end{equation}
\begin{equation}
g^{-1}\partial_{-}g = \left(\begin{matrix}
                               Z_{1} & Z_{2} \\
                               Z_{3} & Z_{4} \\
                             \end{matrix} \right) + H.O \ \ \ \ \ \
                             \ \ \tilde{g}^{-1}\partial_{-}\tilde{g} = \left(\begin{matrix}
                               \tilde{Z}_{1} & \tilde{Z}_{2} \\
                               \tilde{Z}_{3} & \tilde{Z}_{4} \\
                             \end{matrix} \right) + H.O
                             \label{equation3.32}
\end{equation}
\begin{equation}
\begin{array}{cc}
  Z_{1} = \partial_{-}a_{R} \ \ \ \ \ \ \tilde{Z}_{1} = \partial_{-}\tilde{a}_{R} \ \ \ \ \ \ & \ \ \ \ \ \  Z_{4} =
\partial_{-}c_{R} \ \ \ \ \ \ \ \tilde{Z}_{4} =
\partial_{-}\tilde{c}_{R}
\end{array} \notag
\end{equation}
\begin{equation}
Z_{2} = \partial_{-}b_{R} - (a_{L} +
\frac{a_{R}}{2})\partial_{-}b_{R} - (b_{L} +
\frac{b_{R}}{2})\partial_{-}c_{R} + (\partial_{-}a_{R})(b_{L} +
\frac{b_{R}}{2}) + (\partial_{-}b_{R})(c_{L} + \frac{c_{R}}{2})
\notag
\end{equation}
\begin{equation}
\tilde{Z}_{2} = \partial_{-}\tilde{b}_{R} - (\tilde{a}_{L} +
\frac{\tilde{a}_{R}}{2})\partial_{-}\tilde{b}_{R} - (\tilde{b}_{L} +
\frac{\tilde{b}_{R}}{2})\partial_{-}\tilde{c}_{R} +
(\partial_{-}\tilde{a}_{R})(\tilde{b}_{L} + \frac{\tilde{b}_{R}}{2})
+ (\partial_{-}\tilde{b}_{R})(\tilde{c}_{L} +
\frac{\tilde{c}_{R}}{2}) \notag
\end{equation}
\begin{equation}
Z_{3} =  -\partial_{-}b_{R}^{t} + (b_{L}^{t} +
\frac{b_{R}^{t}}{2})\partial_{-}a_{R} + (c_{L} +
\frac{c_{R}}{2})\partial_{-}b_{R}^{t} -
(\partial_{-}b_{R}^{t})(a_{L} + \frac{a_{R}}{2}) -
(\partial_{-}c_{R})(b_{L}^{t} + \frac{b_{R}^{t}}{2}) \notag
\end{equation}
\begin{equation}
\tilde{Z}_{3} =  \partial_{-}\tilde{b}_{R}^{t} - (\tilde{b}_{L}^{t}
+ \frac{\tilde{b}_{R}^{t}}{2})\partial_{-}\tilde{a}_{R} -
(\tilde{c}_{L} +
\frac{\tilde{c}_{R}}{2})\partial_{-}\tilde{b}_{R}^{t} +
(\partial_{-}\tilde{b}_{R}^{t})(\tilde{a}_{L} +
\frac{\tilde{a}_{R}}{2}) +
(\partial_{-}\tilde{c}_{R})(\tilde{b}_{L}^{t} +
\frac{\tilde{b}_{R}^{t}}{2}) \notag
\end{equation}
Obviously equations of motion are satisfied. Since we want to reduce
constraints on the conservation laws and bring the nonlinear
characters of conserved currents into the open we let $e = \sum_{n =
1}^{\infty} \varepsilon^{n}e_{n}$, where e stands for the matrix
components $a$, $b$ and $c$. We may find solutions in the orders of
$\varepsilon$'s. But we need to find transformation matrix $T$ first
and foremost.

\subsubsection{Trivial Case: T = I}

Let us consider first a trivial solution where transformation matrix
is identity. Pseudoduality equations will be
\begin{align}
(\tilde{g_{L}}^{-1}\partial_{+}\tilde{g_{L}})^{i} &= (g_{L}^{-1}\partial_{+}g_{L})^{i} \label{equation3.33} \\
(\tilde{g}^{-1}\partial_{-}\tilde{g})^{i} &= -
(g^{-1}\partial_{-}g)^{i} \label{equation3.34}
\end{align}
Using (\ref{equation3.31}) the first equation (\ref{equation3.33})
leads to
\begin{equation}
\begin{array}{cc}
  \partial_{+}\tilde{a}_{L1} = \partial_{+}a_{L1} \ \ \ \ \ \ \ \ \ \ \ \ \partial_{+}\tilde{a}_{L2} = \partial_{+}a_{L2} \\
  \partial_{+}\tilde{c}_{L1} = \partial_{+}c_{L1} \ \ \ \ \ \ \ \ \ \ \ \ \partial_{+}\tilde{c}_{L2} =
  \partial_{+}c_{L2} \\
  \partial_{+}\tilde{b}_{L1} = \partial_{+}b_{L1} \ \ \ \ \ \ \ \ \ \ \ \ \partial_{+}\tilde{b}_{L1}^{t} =
 -\partial_{+}b_{L1}^{t} \\
  \partial_{+}\tilde{b_{L2}} = \partial_{+}b_{L2} + \frac{1}{2}[A_{L1}(\partial_{+}b_{L1}) + B_{L1}(\partial_{+}c_{L1}) - (\partial_{+}a_{L1})B_{L1} -
  (\partial_{+}b_{L1})C_{L1}] \\
  \partial_{+}\tilde{b}_{L2}^{t} = -\partial_{+}b_{L2}^{t}
  - \frac{1}{2}[B_{L1}^{t}(\partial_{+}a_{L1}) + C_{L1}(\partial_{+}b_{L1}^{t}) - (\partial_{+}b_{L1}^{t})A_{L1} - (\partial_{+}c_{L1})B_{L1}^{t}]
\end{array} \notag
\end{equation}
where we used the solutions of first six equations in the last two
lines as follows
\begin{equation}
\begin{array}{cc}
  \tilde{a}_{L1} = a_{L1} + A_{L1} \ \ \ \ \ \ \ \ \ \ \ \ \tilde{a}_{L2} = a_{L2} + A_{L2} \\
  \tilde{c}_{L1} = c_{L1} + C_{L1} \ \ \ \ \ \ \ \ \ \ \ \ \tilde{c}_{L2} = c_{L2} + C_{L2} \\
  \tilde{b}_{L1} = b_{L1} + B_{L1} \ \ \ \ \ \ \ \ \ \ \ \ \tilde{b}_{L1}^{t} = -b_{L1}^{t} - B_{L1}^{t} \\
  \tilde{b_{L2}} = b_{L2} + \frac{1}{2}(A_{L1}b_{L1} + B_{L1}c_{L1} - a_{L1}B_{L1} -
  b_{L1}C_{L1}) + B_{L2}\\
  \tilde{b}_{L2}^{t} = -b_{L2}^{t} - \frac{1}{2}(B_{L1}^{t}a_{L1} + C_{L1}(\partial_{+}b_{L1}^{t}) - (\partial_{+}b_{L1})A_{L1} -
  c_{L1}B_{L1}^{t}) - B_{L2}^{t}
\end{array} \notag
\end{equation}
where $A_{L1}$, $A_{L2}$, $B_{L1}$, $B_{L2}$, $C_{L1}$ and $C_{L2}$
are constants. Therefore pseudodual left current
(\ref{equation3.31}) up to the order of $\varepsilon^{2}$ in
nonlocal expressions is
\begin{equation}
\tilde{g}_{L}^{-1}\partial_{+}\tilde{g}_{L} = \left(
                                                \begin{array}{cc}
                                                  \tilde{M}_{1} & \tilde{M}_{2} \\
                                                  \tilde{M}_{3} & \tilde{M}_{4} \\
                                                \end{array}
                                              \right) + H.O
                                              \label{equation3.35}
\end{equation}
where we defined the following symbols for the entries of matrix
\begin{align}
  \tilde{M}_{1} &= \varepsilon \partial_{+}\tilde{a}_{L1} + \varepsilon^{2}
  \partial_{+}\tilde{a}_{L2} = \varepsilon \partial_{+}a_{L1} + \varepsilon^{2}
  \partial_{+}a_{L2} \notag\\
  \tilde{M}_{4} &= \varepsilon \partial_{+}\tilde{c}_{L1} + \varepsilon^{2}
  \partial_{+}\tilde{c}_{L2} = \varepsilon \partial_{+}c_{L1} + \varepsilon^{2}
  \partial_{+}c_{L2} \notag
\end{align}
\begin{align}
  \tilde{M}_{2} &= \varepsilon \partial_{+}\tilde{b}_{L1} +
  \varepsilon^{2}[\partial_{+}\tilde{b}_{L2} - \frac{1}{2}(\tilde{a}_{L1}\partial_{+}\tilde{b}_{L1} + \tilde{b}_{L1}\partial_{+}\tilde{c}_{L1} - (\partial_{+}\tilde{a}_{L1})\tilde{b}_{L1} -
  (\partial_{+}\tilde{b}_{L1})\tilde{c}_{L1})] \notag\\ &= \varepsilon
  \partial_{+}b_{L1} + \varepsilon^{2}[\partial_{+}b_{L2} - \frac{1}{2}[a_{L1}(\partial_{+}b_{L1}) + b_{L1}(\partial_{+}c_{L1}) - (\partial_{+}a_{L1})b_{L1} -
  (\partial_{+}b_{L1})c_{L1}]] \notag\\
  \tilde{M}_{3} &= \varepsilon \partial_{+}\tilde{b}_{L1}^{t} +
  \varepsilon^{2}[\partial_{+}\tilde{b}_{L2}^{t} - \frac{1}{2}[\tilde{b}_{L1}^{t}(\partial_{+}\tilde{a}_{L1}) + \tilde{c}_{L1}(\partial_{+}\tilde{b}_{L1}^{t}) - (\partial_{+}\tilde{b}_{L1}^{t})\tilde{a}_{L1} -
  (\partial_{+}\tilde{c}_{L1})\tilde{b}_{L1}^{t}]] \notag\\
  &= - \varepsilon \partial_{+}b_{L1}^{t} - \varepsilon^{2}[\partial_{+}b_{L2}^{t} - \frac{1}{2}[b_{L1}^{t}(\partial_{+}a_{L1}) + c_{L1}(\partial_{+}b_{L1}^{t}) - (\partial_{+}b_{L1}^{t})a_{L1} -
  (\partial_{+}c_{L1})b_{L1}^{t}]] \notag
\end{align}
Obviously this current is conserved. To find right current we use
$2^{nd}$ pseudoduality equation (\ref{equation3.34}) and we find the
following expressions up to the order of $\varepsilon^{2}$
\begin{equation}
\begin{array}{cc}
  \partial_{-}\tilde{a}_{R1} = - \partial_{-}a_{R1} \ \ \ \ \ \ \ \ \partial_{-}\tilde{a}_{R2} = - \partial_{-}a_{R2} \\
  \partial_{-}\tilde{c}_{R1} = - \partial_{-}c_{R1} \ \ \ \ \ \ \ \ \partial_{-}\tilde{c}_{R2}
  = - \partial_{-}c_{R2} \\
  \partial_{-}\tilde{b}_{R1} = - \partial_{-}b_{R1} \ \ \ \ \ \ \ \ \partial_{-}\tilde{b}_{R1}^{t} =
 \partial_{-}b_{R1}^{t} \\
  \partial_{-}\tilde{b_{R2}} = - \partial_{-}b_{R2} + (a_{R1} - A_{L1} + \frac{A_{R1}}{2})(\partial_{-}b_{R1}) + (b_{R1} - B_{L1} + \frac{B_{R1}}{2})(\partial_{-}c_{R1})\\ - (\partial_{-}a_{R1})(b_{R1} - B_{L1} + \frac{B_{R1}}{2}) -
  (\partial_{-}b_{R1})(c_{R1} - C_{L1} + \frac{C_{R1}}{2}) \\
  \partial_{-}\tilde{b}_{R2}^{t} = \partial_{-}b_{R2}^{t}
  - (- B_{L1}^{t} + b_{R1}^{t} + \frac{B_{R1}^{t}}{2})(\partial_{-}a_{R1}) - (-C_{L1} + c_{R1} + \frac{C_{R1}}{2})(\partial_{-}b_{R1}^{t})\\ + (\partial_{-}b_{R1}^{t})(- A_{L1} + a_{R1} + \frac{A_{R1}}{2}) + (\partial_{-}c_{R1})(- B_{L1}^{t} + b_{R1}^{t} +
  \frac{B_{R1}^{t}}{2})
\end{array} \notag
\end{equation}
where we used the solution of first six equations in the last two
equations as
\begin{equation}
\begin{array}{cc}
  \tilde{a}_{R1} = - a_{R1} - A_{R1} \ \ \ \ \ \ \tilde{a}_{R2} = - a_{R2} - A_{R2} \\
  \tilde{c}_{R1} = - c_{R1} - C_{R1} \ \ \ \ \ \ \tilde{c}_{R2} = - c_{R2} - C_{R2} \\
  \tilde{b}_{R1} = - b_{R1} - B_{R1} \ \ \ \ \ \ \tilde{b}_{R1}^{t} = b_{R1}^{t} + B_{R1}^{t} \\
  \end{array} \notag
\end{equation}
where $A_{R1}$, $A_{R2}$, $B_{R1}$, $C_{R1}$ and $C_{R2}$ are
constants. A brief computation yields the following expression for
the right current
\begin{equation}
(\partial_{-}\tilde{g}_{R})\tilde{g}_{R}^{-1} = \left(
                                                \begin{array}{cc}
                                                  \tilde{N}_{1} & \tilde{N}_{2} \\
                                                  \tilde{N}_{3} & \tilde{N}_{4} \\
                                                \end{array}
                                              \right) + H.O
                                              \label{equation3.36}
\end{equation}
\begin{equation}
\begin{array}{cc}
  \tilde{N}_{1} = \varepsilon \partial_{-}\tilde{a}_{R1} + \varepsilon^{2}
  \partial_{-}\tilde{a}_{R2} = - \varepsilon \partial_{-}a_{R1} - \varepsilon^{2}
  \partial_{-}a_{R2}
  \\
  \tilde{N}_{4} = \varepsilon \partial_{-}\tilde{c}_{R1} + \varepsilon^{2}
  \partial_{-}\tilde{c}_{R2} = - \varepsilon \partial_{-}c_{R1} - \varepsilon^{2}
  \partial_{-}c_{R2} \\
  \tilde{N}_{2} = \varepsilon \partial_{-}\tilde{b}_{R1} +
  \varepsilon^{2}[\partial_{-}\tilde{b}_{R2} + \frac{1}{2}(\tilde{a}_{R1}\partial_{-}\tilde{b}_{R1} + \tilde{b}_{R1}\partial_{-}\tilde{c}_{R1} - (\partial_{-}\tilde{a}_{R1})\tilde{b}_{R1} -
  (\partial_{-}\tilde{b}_{R1})\tilde{c}_{R1})] \\ = - \varepsilon
  \partial_{-}b_{R1} + \varepsilon^{2}[- \partial_{-}b_{R2} +(\frac{3}{2}a_{R1} + A_{R1} - A_{L1})(\partial_{-}b_{R1}) + (\frac{3}{2}b_{R1} + B_{R1} - B_{L1})(\partial_{-}c_{R1})\\ - (\partial_{-}a_{R1})(\frac{3}{2}b_{R1} + B_{R1} - B_{L1}) -
  (\partial_{-}b_{R1})(\frac{3}{2}c_{R1} + C_{R1} - C_{L1})]\\
  \tilde{N}_{3} = \varepsilon \partial_{-}\tilde{b}_{R1}^{t} +
  \varepsilon^{2}[\partial_{-}\tilde{b}_{R2}^{t} + \frac{1}{2}[\tilde{b}_{R1}^{t}(\partial_{-}\tilde{a}_{R1}) + \tilde{c}_{R1}(\partial_{-}\tilde{b}_{R1}^{t}) - (\partial_{-}\tilde{b}_{R1}^{t})\tilde{a}_{R1} -
  (\partial_{-}\tilde{c}_{R1})\tilde{b}_{R1}^{t}]] \\
  = \varepsilon \partial_{-}b_{R1}^{t} + \varepsilon^{2}[\partial_{-}b_{R2}^{t} - (\frac{3}{2}b_{R1}^{t} + B_{R1}^{t} -B_{L1}^{t})(\partial_{-}a_{R1}) - (\frac{3}{2}c_{R1} + C_{R1} - C_{L1})(\partial_{-}b_{R1}^{t})\\ + (\partial_{-}b_{R1}^{t})(\frac{3}{2}a_{R1} + A_{R1} - A_{L1}) + (\partial_{-}c_{R1})(\frac{3}{2}b_{R1}^{t} + B_{R1}^{t} - B_{L1}^{t})]
\end{array} \notag
\end{equation}

We see that this current is also conserved.

\subsubsection{Nontrivial Case: General T}

In this case we use the general expression (\ref{equation3.19}) of
transformation matrix T. Pseudoduality equations are given by
(\ref{equation3.4}) and (\ref{equation3.5}), and gave us the
equations (\ref{equation3.22}) and (\ref{equation3.23}) which can be
written as
\begin{align}
\partial_{+}\tilde{a}_{L1} &= \partial_{+}a_{L1} \ \ \ \ \
\partial_{+}\tilde{b}_{L1} = \partial_{+}b_{L1} \ \ \ \ \ \partial_{+}\tilde{b}_{L1}^{t} = -\partial_{+}b_{L1}^{t} \ \ \ \ \
\partial_{+}\tilde{c}_{L1} = \partial_{+}c_{L1} \notag\\
\partial_{-}\tilde{a}_{R1} &= -\partial_{-}a_{R1} \ \ \ \ \
\partial_{-}\tilde{b}_{R1} = -\partial_{-}b_{R1} \ \ \ \ \ \partial_{-}\tilde{b}_{R1}^{t} = \partial_{-}b_{R1}^{t} \ \ \ \ \
\partial_{-}\tilde{c}_{R1} = -\partial_{-}c_{R1} \notag\\
\partial_{+}\tilde{a}_{L2} &= \partial_{+}a_{L2} \ \ \ \ \
\partial_{+}\tilde{c}_{L2} = \partial_{+}c_{L2} \ \ \ \ \ \partial_{-}\tilde{a}_{R2} = -\partial_{-}a_{R2} \ \ \ \ \
\partial_{-}\tilde{c}_{R2} = -\partial_{-}c_{R2} \notag\\
\partial_{+}\tilde{b}_{L2} &= \partial_{+}b_{L2} -\frac{1}{2}[A_{L1}(\partial_{+}b_{L1}) + B_{L1}(\partial_{+}c_{L1}) - (\partial_{+}a_{L1})B_{L1} -
(\partial_{+}b_{L1})C_{L1}] \notag\\
\partial_{+}\tilde{b}_{L2}^{t} &= -\partial_{+}b_{L2}^{t} +
\frac{1}{2}[B_{L1}^{t}(\partial_{+}a_{L1}) +
C_{L1}(\partial_{+}b_{L1}^{t}) - (\partial_{+}b_{L1}^{t})A_{L1} -
(\partial_{+}c_{L1})B_{L1}^{t}] \notag\\
\partial_{-}\tilde{b}_{R2} &= -\partial_{-}b_{R2} + (a_{R1} +
\frac{A_{R1}}{2})(\partial_{-}b_{R1}) + (b_{R1} +
\frac{B_{R1}}{2})(\partial_{-}c_{R1}) \notag\\ &-
(\partial_{-}a_{R1})(b_{R1} + \frac{B_{R1}}{2}) -
(\partial_{-}b_{R1})(c_{R1} + \frac{C_{R1}}{2}) \notag\\
\partial_{-}\tilde{b}_{R2}^{t} &= \partial_{-}b_{R2}^{t} - (b_{R1}^{t} +
\frac{B_{R1}^{t}}{2})(\partial_{-}a_{R1}) - (c_{R1} +
\frac{C_{R1}}{2})(\partial_{-}b_{R1}^{t}) \notag\\ &+
(\partial_{-}b_{R1}^{t})(a_{R1} + \frac{A_{R1}}{2}) +
(\partial_{-}c_{R1})(b_{R1}^{t} + \frac{B_{R1}^{t}}{2}) \notag
\end{align}
where we used the solutions of first three lines for the last four
expressions. Solutions of these equations are
\begin{align}
\tilde{a}_{L1} = a_{L1} + A_{L1} \ \ \ \ \ \tilde{b}_{L1} = b_{L1} +
B_{L1} \ \ \ \ \ \tilde{b}_{L1}^{t} = -b_{L1}^{t} - B_{L1}^{t} \notag\\
\tilde{c}_{L1} = c_{L1} + C_{L1} \ \ \ \ \ \tilde{a}_{R1} = -a_{R1}
- A_{R1}
\ \ \ \ \ \tilde{b}_{R1} = -b_{R1} - B_{R1} \notag\\
\tilde{b}_{R1}^{t} = b_{R1}^{t} + B_{R1}^{t} \ \ \ \ \
\tilde{c}_{R1} =
-c_{R1} - C_{R1} \ \ \ \ \ \tilde{a}_{L2} = a_{L2} + A_{L2} \notag\\
\tilde{c}_{L2} = c_{L2} + C_{L2} \ \ \ \ \ \tilde{a}_{R2} = -a_{R2}
- A_{R2} \ \ \ \
\ \tilde{c}_{R2} = -c_{R2} - C_{R2} \notag\\
\tilde{b}_{L2} = b_{L2} + B_{L2} - \frac{1}{2}[A_{L1}b_{L1} +
B_{L1}c_{L1} - a_{L1}B_{L1} - b_{L1}C_{L1}] \notag\\
\tilde{b}_{L2}^{t} = - b_{L2}^{t} - B_{L2}^{t} +
\frac{1}{2}[B_{L1}^{t}a_{L1} + C_{L1}b_{L1}^{t} - b_{L1}^{t}A_{L1} -
c_{L1}B_{L1}^{t}] \notag
\end{align}
where $A_{L1}$, $A_{R1}$, $B_{L1}$, $B_{R1}$, $C_{L1}$, $C_{R1}$,
and $B_{L2}$ are constants. We did not find solutions of
$\tilde{b}_{R2}$ and $\tilde{b}_{R2}^{t}$ because of their
complicated forms and no need to use them. Hence pseudodual left
current (\ref{equation3.24}) will be
\begin{align}
\tilde{J}_{+}^{(L)} &= \tilde{g}^{-1}\partial_{+}\tilde{g} =
\varepsilon \partial_{+}\tilde{y}_{L1} +
\varepsilon^{2}\{\partial_{+}\tilde{y}_{L2} -
\frac{1}{2}[\tilde{y}_{L1},
\partial_{+}\tilde{y}_{L1}]_{\tilde{G}}\} + H.O. \notag\\
&= \left(
                                                              \begin{array}{cc}
                                                                \tilde{M}_{1} & \tilde{M}_{2} \\
                                                                \tilde{M}_{3} & \tilde{M}_{4} \\
                                                              \end{array}
                                                            \right)
+ H.O. \label{equation3.37}
\end{align}
where
\begin{align}
\tilde{M}_{1} &= \varepsilon \partial_{+}\tilde{a}_{L1} +
\varepsilon^{2}\partial_{+}\tilde{a}_{L2} = \varepsilon
\partial_{+}a_{L1} +
\varepsilon^{2}\partial_{+}a_{L2} \notag\\
\tilde{M}_{4} &= \varepsilon \partial_{+}\tilde{c}_{L1} +
\varepsilon^{2}\partial_{+}\tilde{c}_{L2} = \varepsilon
\partial_{+}c_{L1} +
\varepsilon^{2}\partial_{+}c_{L2} \notag\\
\tilde{M}_{2} &= \varepsilon \partial_{+}\tilde{b}_{L1} +
\varepsilon^{2}[\partial_{+}\tilde{b}_{L2} -
\frac{1}{2}\{\tilde{a}_{L1}(\partial_{+}\tilde{b}_{L1}) +
\tilde{b}_{L1}(\partial_{+}\tilde{c}_{L1}) -
(\partial_{+}\tilde{a}_{L1})\tilde{b}_{L1} -
(\partial_{+}\tilde{b}_{L1})\tilde{c}_{L1}\}] \notag\\
&= \varepsilon\partial_{+}b_{L1} +
\varepsilon^{2}[\partial_{+}b_{L2} -
\frac{1}{2}\{a_{L1}(\partial_{+}b_{L1}) + b_{L1}(\partial_{+}c_{L1})
- (\partial_{+}a_{L1})b_{L1} - (\partial_{+}b_{L1})c_{L1}\}]
\notag\\
\tilde{M}_{3} &= \varepsilon \partial_{+}\tilde{b}_{L1}^{t} +
\varepsilon^{2}[\partial_{+}\tilde{b}_{L2}^{t} -
\frac{1}{2}\{\tilde{b}_{L1}^{t}(\partial_{+}\tilde{a}_{L1}) +
\tilde{c}_{L1}(\partial_{+}\tilde{b}_{L1}^{t}) -
(\partial_{+}\tilde{b}_{L1}^{t})\tilde{a}_{L1} -
(\partial_{+}\tilde{c}_{L1})\tilde{b}_{L1}^{t}\}] \notag\\
&= - \varepsilon \partial_{+}b_{L1}^{t} -
\varepsilon^{2}[\partial_{+}b_{L2}^{t} - (\frac{b_{L1}^{t}}{2} +
B_{L1}^{t})(\partial_{+}a_{L1}) - (\frac{c_{L1}}{2} +
C_{L1})(\partial_{+}b_{L1}^{t})\notag\\ &+
(\partial_{+}b_{L1}^{t})(\frac{a_{L1}}{2} + A_{L1}) +
(\partial_{+}c_{L1})(\frac{b_{L1}^{t}}{2} + B_{L1}^{t})] \notag
\end{align}

Pseudodual right current (\ref{equation3.25}) can be constructed as
follows
\begin{align}
\tilde{J}_{-}^{(R)} &= (\partial_{-}\tilde{g})\tilde{g}^{-1} =
\varepsilon \partial_{-}\tilde{y}_{R1} +
\varepsilon^{2}\{\partial_{-}\tilde{y}_{R2} +
\frac{1}{2}[\tilde{y}_{R1},
\partial_{-}\tilde{y}_{R1}]_{\tilde{G}}\} + H.O. \notag\\
&= \left(
     \begin{array}{cc}
       \tilde{N}_{1} & \tilde{N}_{2} \\
       \tilde{N}_{3} & \tilde{N}_{4} \\
     \end{array}
   \right) + H.O. \label{equation3.38}
\end{align}
where
\begin{align}
\tilde{N}_{1} &= \varepsilon \partial_{-}\tilde{a}_{R1} +
\varepsilon^{2} \partial_{-}\tilde{a}_{R2} = - \varepsilon
\partial_{-}a_{R1} - \varepsilon^{2}
\partial_{-}a_{R2} \notag\\
\tilde{N}_{4} &= \varepsilon \partial_{-}\tilde{c}_{R1} +
\varepsilon^{2} \partial_{-}\tilde{c}_{R2} = - \varepsilon
\partial_{-}c_{R1} - \varepsilon^{2}
\partial_{-}c_{R2} \notag\\
\tilde{N}_{2} &= \varepsilon \partial_{-}\tilde{b}_{R1} +
\varepsilon^{2}\{\partial_{-}\tilde{b}_{R2} +
\frac{1}{2}[\tilde{a}_{R1}(\partial_{-}\tilde{b}_{R1}) +
\tilde{b}_{R1}(\partial_{-}\tilde{c}_{R1}) -
(\partial_{-}\tilde{a}_{R1})\tilde{b}_{R1} -
(\partial_{-}\tilde{b}_{R1})\tilde{c}_{R1}]\} \notag\\
&= -\varepsilon \partial_{-}b_{R1} -
\varepsilon^{2}\{\partial_{-}b_{R2} -(\frac{3a_{R1}}{2} +
A_{R1})(\partial_{-}b_{R1}) - (\frac{3b_{R1}}{2} +
B_{R1})(\partial_{-}c_{R1}) \notag\\ &+
(\partial_{-}a_{R1})(\frac{3b_{R1}}{2} + B_{R1}) +
(\partial_{-}b_{R1})(\frac{3c_{R1}}{2} + C_{R1})\}
\notag\\
\tilde{N}_{3} &= \varepsilon
\partial_{-}\tilde{b}_{R1}^{t} +
\varepsilon^{2}\{\partial_{-}\tilde{b}_{R2}^{t} +
\frac{1}{2}[\tilde{b}_{R1}^{t}(\partial_{-}\tilde{a}_{R1}) +
\tilde{c}_{R1}(\partial_{-}\tilde{b}_{R1}^{t}) -
(\partial_{-}\tilde{b}_{R1}^{t})\tilde{a}_{R1} -
(\partial_{-}\tilde{c}_{R1})\tilde{b}_{R1}^{t}]\} \notag\\
&=\varepsilon\partial_{-}b_{R1}^{t} +
\varepsilon^{2}\{\partial_{-}b_{R2}^{t} - (\frac{3b_{R1}^{t}}{2} +
B_{R1}^{t})(\partial_{-}a_{R1}) - (\frac{3c_{R1}}{2} +
C_{R1})(\partial_{-}b_{R1}^{t}) \notag\\ &+
(\partial_{-}b_{R1}^{t})(\frac{3a_{R1}}{2} + A_{R1}) +
(\partial_{-}c_{R1})(\frac{3b_{R1}^{t}}{2} + B_{R1}^{t})\} \notag
\end{align}

It is apparent that these currents are conserved.

\section{Cartan Decomposition of Symmetric Spaces} \label{sec:CDSS}

We saw in the above example that symmetric spaces can be decomposed
into two pieces, one piece remains invariant under transformation T
though the other piece is transformed in such a way that it behaves
like a new symmetric space. Let $\pi$ be the projection $G
\longrightarrow M$, sending each $g \in G$ to submersion $M$. We see
that $M$ is symmetric space after invariant parts of $G$ are
eliminated.

Let $H$ be a closed subgroup of a connected Lie group G, and
$\sigma$ be an involutive automorphism of G such that $F_{0} \subset
H \subset F = Fix(\sigma)$. Symmetric space $M$ is the coset space
$M = G / H$. If $\textbf{g}$ is the Lie algebra of $G$, $\textbf{h}$
is the Lie algebra of $H$, and $\textbf{m}$ is the Lie subspace
\footnote{\textbf{m} is called as the Lie subspace for $M$, not Lie
Algebra \cite{oneill}.} for $M$, then $\textbf{g} = \textbf{m}
\oplus \textbf{h}$, where $\textbf{h}$ is closed under brackets
while $\textbf{m}$ is $Ad(H)$-invariant subspace of $\textbf{g}$,
i.e, $Ad_{h}(\textbf{m}) \subset \textbf{m}$ for all $h \in H$. If
$X \in \textbf{g}$, then $X = X_{h} + X_{m}$, where $X_{h} \in
\textbf{h}$, and $X_{m} \in \textbf{m}$. The involutive automorphism
$d\sigma$ is such that $d\sigma(X_{h}) = X_{h}$ and $d\sigma(X_{m})
= -X_{m}$. Bracket relations for the symmetric space are defined by
\begin{equation}
[\textbf{h}, \textbf{h}] \subset \textbf{h},\ \ \ \ \ [\textbf{h},
\textbf{m}] \subset \textbf{m},\ \ \ \ \ [\textbf{m}, \textbf{m}]
\subset \textbf{h} \label{equation3.39}
\end{equation}

The currents $J_{+}^{(L)} = g^{-1}\partial_{+}g$ and $J_{-}^{(R)} =
(\partial_{-}g)g^{-1}$ on $\textbf{g}$ can be split into the
currents $J_{m}^{(L)} = g^{-1}D_{+}g$ and $J_{m}^{(R)} =
(D_{-}g)g^{-1}$ on $\textbf{m}$ and $J_{h}^{(L)} = A_{+}$ and
$J_{h}^{(R)} = gA_{-}g^{-1}$ on $\textbf{h}$, where $D_{\pm}$ is the
covariant derivative acting on $\textbf{m}$, and $A_{\pm}$ is the
gauge field defined on $\textbf{h}$.

If one defines indices $i, j, k, ...$ for the space elements of
$\textbf{g}$, indices $a, b, c, ...$ for the space elements of
$\textbf{h}$, and indices $\alpha, \beta, \gamma, ...$ for the space
elements of $\textbf{m}$, then (\ref{equation3.39}) allows only
structure constants $f_{bc}^{a}$, $f_{a \beta}^{\alpha}$, $f_{\beta
a}^{\alpha}$, and $f_{\alpha\beta}^{a}$. The other structure
constants vanish. This leads to the following equations of motion,
\begin{align}
k_{+} &= g^{-1}D_{+}g \ \ \ \ \ \Longrightarrow \ \ \ \ \ D_{-}k_{+}
= 0 \label{equation3.40} \\
k_{-} &= g^{-1}D_{-}g \ \ \ \ \ \Longrightarrow \ \ \ \ \ D_{+}k_{-}
= [k_{-}, A_{+}] + [A_{-}, k_{+}] \label{equation3.41} \\
A_{+} &= g^{-1}D_{+}^{'}g \ \ \ \ \ \Longrightarrow \ \ \ \ \
D_{-}^{'}A_{+} = 0 \label{equation3.42} \\
A_{-} &= g^{-1}D_{-}^{'}g \ \ \ \ \ \Longrightarrow \ \ \ \ \
D_{+}^{'}A_{-} = [A_{-}, A_{+}] + [k_{-}, k_{+}]
\label{equation3.43}
\end{align}
where $k_{\pm}$($A_{\pm}$) belongs to $\textbf{m}$($\textbf{h}$),
and $D$($D^{'}$) is the covariant derivative acting on
$\textbf{m}$($\textbf{h}$).

It is natural to write down the Pseudoduality equations
(\ref{equation3.4}) and (\ref{equation3.5}) in the most general
split form on two spaces $\textbf{m}$ and $\textbf{h}$ as follows
\begin{equation} \label{equation3.44}
  \begin{array}{cc}
    \tilde{k}_{+}^{\alpha} = T_{\beta}^{\alpha} k_{+}^{\beta} + T_{a}^{\alpha} A_{+}^{a} \ \ \ \  & \ \ \ \ \tilde{A}_{+}^{a} = T_{b}^{a} A_{+}^{b} + T_{\alpha}^{a} k_{+}^{\alpha} \\
    \tilde{k}_{-}^{\alpha} = - T_{\beta}^{\alpha} k_{-}^{\beta} - T_{a}^{\alpha} A_{-}^{a} \ \ \ \  & \ \ \ \  \tilde{A}_{-}^{a} = - T_{b}^{a} A_{-}^{b} - T_{\alpha}^{a} k_{-}^{\alpha} \\
  \end{array}
\end{equation}
where
\begin{equation} \label{equation3.45}
g^{-1}\partial_{+}g = \left(
                          \begin{array}{c}
                            k_{+} \\
                            A_{+} \\
                          \end{array}
                        \right) \ \ \begin{array}{c}
                                      $on$ \ \ $\textbf{m}$-$space$ \\
                                      $on$ \ \ $\textbf{h}$-$space$
                                    \end{array} \ \ \
\end{equation}
\begin{equation} \label{equation3.46}
g^{-1}\partial_{-}g = \left(
                           \begin{array}{c}
                             k_{-} \\
                             A_{-} \\
                           \end{array}
                         \right) \ \ \begin{array}{c}
                                       $on$ \ \ $\textbf{m}$-$space$  \\
                                       $on$ \ \ $\textbf{h}$-$space$
                                     \end{array}
\end{equation}
and
\begin{align} \label{equation3.47}
T = \left(
      \begin{array}{cc}
        T_{\beta}^{\alpha} & T_{a}^{\alpha} \\
        T_{\beta}^{a} & T_{b}^{a} \\
      \end{array}
    \right) \begin{array}{c}
              $on$ \ \ $\textbf{m}$-$space$ \\
              $on$ \ \ $\textbf{h}$-$space$
            \end{array}
\end{align}

Apparently $T_{a}^{\alpha}$ and $T_{\beta}^{a}$ represent the mixing
components of the isometry preserving map $T$. Before considering
this most general pseudoduality relations which lead to mixed
expressions it is worth to analyze pseudoduality equations between
pure symmetric spaces and their counter $H$-spaces without mixing
parts.

\subsection{Non-Mixing Pseudoduality} \label{sec:NMP}

We set the mixing components $T_{a}^{\alpha}$ and $T_{\beta}^{a}$ in
equation (\ref{equation3.44}) equal to zero, and consider the
pseudoduality equations on $\textbf{m}$ and $\textbf{h}$-spaces as
follows
\begin{align}
\tilde{k}_{\pm}^{\alpha} = \pm T_{\beta}^{\alpha} k_{\pm}^{\beta} \label{equation3.48}\\
\tilde{A}_{\pm}^{a} = \pm T_{b}^{a}A_{\pm}^{b} \label{equation3.49}
\end{align}
When we take $D_{-}$ of (\ref{equation3.48}), and $D_{-}^{'}$ of
(\ref{equation3.49}) (`+' equations only) followed by the equations
of motion (\ref{equation3.40}) and (\ref{equation3.42}) we obtain
the result that both $T_{\beta}^{\alpha}$ and $T_{b}^{a}$ depend
only on $\sigma^{+}$. Now let us take $D_{+}$ of `$-$' equation in
(\ref{equation3.48}), and use (\ref{equation3.41}) to get
\begin{equation}
[\tilde{k}_{-}, \tilde{A}_{+}]^{\alpha} + [\tilde{A}_{-},
\tilde{k}_{+}]^{\alpha} = - (D_{+} T_{\beta}^{\alpha}) k_{-}^{\beta}
- T_{\beta}^{\alpha} [k_{-}, A_{+}]^{\beta} - T_{\beta}^{\alpha}
[A_{-}, k_{+}]^{\beta} \label{equation3.50}
\end{equation}
Since $k_{-}$ and $A_{-}$ can be treated independently, this
equation can be split into the following equations
\begin{align}
\tilde{f}_{\beta a}^{\alpha} \tilde{k}_{+}^{\beta} T_{c}^{a} &=
T_{\beta}^{\alpha}
f_{\lambda c}^{\beta} k_{+}^{\lambda} \label{equation3.51}\\
\tilde{f}_{a \beta}^{\alpha} \tilde{A}_{+}^{a} T_{\lambda}^{\beta}
&= - D_{+} T_{\lambda}^{\alpha} + T_{\beta}^{\alpha} f_{a
\lambda}^{\beta} A_{+}^{a} \label{equation3.52}
\end{align}
First equation (\ref{equation3.51}) gives us a relation between
structure constants, $\tilde{f}_{\beta a}^{\alpha}
T_{\lambda}^{\beta} T_{c}^{a} = T_{\beta}^{\alpha} f_{\lambda
c}^{\beta}$, which leads second equation to yield
$D_{+}T_{\lambda}^{\alpha} = 0$. Therefore we conclude that
$T_{\beta}^{\alpha}$ has to be a constant, and we choose it to be
identity. Similarly we take $D_{+}^{'}$ of `$-$' equation in
(\ref{equation3.49}), and use (\ref{equation3.43}) to get
\begin{equation}
[\tilde{A}_{-}, \tilde{A}_{+}]^{a} + [\tilde{k}_{-},
\tilde{k}_{+}]^{a} = - (D_{+}^{'} T_{b}^{a}) A_{-}^{b} - T_{b}^{a}
[A_{-}, A_{+}]^{b} - T_{b}^{a} [k_{-}, k_{+}]^{b}
\label{equation3.53}
\end{equation}
This equation yields the following results
\begin{align}
\tilde{f}_{\alpha \beta}^{a} \tilde{k}_{+}^{\alpha}
T_{\lambda}^{\beta} &= T_{b}^{a}
f_{\beta \lambda}^{b} k_{+}^{\beta} \label{equation3.54}\\
\tilde{f}_{b c}^{a} \tilde{A}_{+}^{b} T_{d}^{c} &= - D_{+}^{'}
T_{d}^{a} + T_{b}^{a} f_{c d}^{b} A_{+}^{c} \label{equation3.55}
\end{align}
First equation (\ref{equation3.54}) verifies the result above up to
the permutation of indices, $\tilde{f}_{\alpha \beta}^{a}
T_{\nu}^{\alpha} T_{\lambda}^{\beta} = T_{b}^{a} f_{\nu
\lambda}^{b}$. Second equation (\ref{equation3.55}) produces the
following solution
\begin{equation}
T_{b}^{a} = T_{b}^{a} (0) + (f_{c b}^{a} - \tilde{f}_{c b}^{a})
\int_{0}^{\sigma^{+}}  A_{+}^{c} D^{'} \sigma^{'+} + H.O.
\label{equation3.56}
\end{equation}
where we choose $T_{b}^{a} (0)$ to be identity. It is easy to see
that these equations yield the following bracket relations
\begin{align}
[\tilde{k}_{+}, \tilde{A}_{-}]^{\alpha} &= - T_{\beta}^{\alpha}
[k_{+}, A_{-}]^{\beta} \label{equation3.57}\\
[\tilde{k}_{-}, \tilde{A}_{+}]^{\alpha} &= - T_{\beta}^{\alpha}
[k_{-}, A_{+}]^{\beta} \label{equation3.58}\\
[\tilde{k}_{+}, \tilde{k}_{-}]^{a} &= - T_{b}^{a} [k_{+}, k_{-}]^{b} \label{equation3.59}\\
[\tilde{A}_{+}, \tilde{A}_{-}]^{a} &= - T_{b}^{a} [A_{+}, A_{-}]^{b}
+ (D_{+}^{'} T_{b}^{a}) A_{-}^{b} \label{equation3.60}
\end{align}
that verifies the equations of motion on pseudodual space as pointed
out above, $D_{+} \tilde{k}_{-}^{\alpha} = - T_{\beta}^{\alpha}
D_{+} k_{-}^{\beta}$ and $D_{+}\tilde{A}_{-}^{a} = - T_{b}^{a}
D_{+}^{'} A_{-}^{b} - (D_{+}^{'} T_{b}^{a}) A_{-}^{b}$. We notice
that if $H$ and $\tilde{H}$ are the same for both manifolds, i.e.,
$f_{b c}^{a} = \tilde{f}_{b c}^{a}$, then $T_{b}^{a}$ reduces to
identity, and we recover the flat space pseudoduality relations on
two manifolds. One can easily construct nonlocal field expressions
using above solutions, which are
\begin{align}
\tilde{k}_{\pm} &= \pm k_{\pm} \label{equation3.61}\\
\tilde{A}_{\pm} &= \pm A_{\pm} \pm \int_{0}^{\sigma^{+}} ([A_{+}
(\sigma^{'+}), A_{\pm} (\sigma^{+})]_{H} - [A_{+} (\sigma^{'+}),
A_{\pm} (\sigma^{+})]_{\tilde{H}}) D^{'}\sigma^{'+} + H.O.
\label{equation3.62}
\end{align}
One may readily construct nonlocal expressions of the conserved
pseudodual currents by means of these fields and following the
method in section 2 (\ref{sec:WZWCC}).

\subsection{Mixing Pseudoduality} \label{MP}

We now consider mixing of $\textbf{m}$ and $\textbf{h}$-spaces in
pseudodual expressions. Pseudoduality equations can be written as in
(\ref{equation3.44}). We take $\partial_{-}$ of first equation on
$\textbf{m}$-space (\ref{equation3.44}), and obtain
\begin{equation} \label{equation3.63}
(\partial_{-}T_{\beta}^{\alpha})k_{+}^{\beta} +
(\partial_{-}T_{a}^{\alpha}) A_{+}^{a} = 0
\end{equation}
since $\textbf{m}$ and $\textbf{h}$-spaces are independent, we get
$\partial_{-}T_{\beta}^{\alpha} = \partial_{-}T_{a}^{\alpha} = 0$,
so $T_{\beta}^{\alpha}$ and $T_{a}^{\alpha}$ don't depend on
$\sigma^{-}$. Now we take $\partial_{+}$ of second equation on
$\textbf{m}$-space (\ref{equation3.44}) and see that
\begin{align} \label{equation3.64}
[\tilde{k}_{-}, \tilde{A}_{+}]^{\alpha} + [\tilde{A}_{-},
\tilde{k}_{+}]^{\alpha} = &- (\partial_{+} T_{\beta}^{\alpha})
k_{-}^{\beta} - T_{\beta}^{\alpha} [k_{-}, A_{+}]^{\beta} -
T_{\beta}^{\alpha} [A_{-}, k_{+}]^{\beta} \notag\\ &- (\partial_{+}
T_{a}^{\alpha}) A_{-}^{a} - T_{a}^{\alpha} [A_{-}, A_{+}]^{a} -
T_{a}^{\alpha} [k_{-}, k_{+}]^{a}
\end{align}
We substitute the expressions for $\tilde{k}_{-}$ and
$\tilde{A}_{-}$ into this equation, and compare the coefficients of
$k_{-}$ and $A_{-}$ to get the following expressions
\begin{align}
\partial_{+} T_{\lambda}^{\alpha} = [f_{b \lambda}^{\beta} T_{\beta}^{\alpha} - \tilde{f}_{a \beta}^{\alpha} (T_{b}^{a} T_{\lambda}^{\beta} - T_{b}^{\beta}
T_{\lambda}^{a})]A_{+}^{b} + [f_{\beta \lambda}^{a} T_{a}^{\alpha} -
\tilde{f}_{a \nu}^{\alpha} (T_{\beta}^{a} T_{\lambda}^{\nu} -
T_{\beta}^{\nu} T_{\lambda}^{a})] k_{+}^{\beta} \label{equation3.65}\\
\partial_{+} T_{b}^{\alpha} = [f_{\beta b}^{\nu} T_{\nu}^{\alpha} - \tilde{f}_{a \nu}^{\alpha} (T_{\beta}^{a} T_{b}^{\nu} - T_{\beta}^{\nu}
T_{b}^{a})]k_{+}^{\beta} + [f_{c b}^{a} T_{a}^{\alpha} -
\tilde{f}_{\beta a}^{\alpha} (T_{c}^{\beta} T_{b}^{a} - T_{c}^{a}
T_{b}^{\beta})] A_{+}^{c} \label{equation3.66}
\end{align}
Since we only need to find currents up to the second order terms, it
suffices to find mapping tensors using only initial values
\begin{align}
T_{\lambda}^{\alpha} (\sigma^{+}) = &T_{\lambda}^{\alpha} (0) +
(f_{b \lambda}^{\alpha} - \tilde{f}_{b \lambda}^{\alpha} +
\tilde{f}_{a \beta}^{\alpha} T_{b}^{\beta} (0) T_{\lambda}^{a} (0))
\int_{0}^{\sigma^{+}} A_{+}^{b} D^{'} \sigma^{'+} \label{equation3.67}\\
&+ (f_{\beta \lambda}^{a} T_{a}^{\alpha} (0) - \tilde{f}_{a
\lambda}^{\alpha} T_{\beta}^{a} (0) + \tilde{f}_{a \beta}^{\alpha}
T_{\lambda}^{a} (0)) \int_{0}^{\sigma^{+}} k_{+}^{\beta} D
\sigma^{'+} + H.O. \notag
\end{align}
\begin{align}
T_{b}^{\alpha} (\sigma^{+}) = &T_{b}^{\alpha} (0) + (f_{\beta
b}^{\alpha} + \tilde{f}_{b \beta}^{\alpha} - \tilde{f}_{a
\nu}^{\alpha} T_{\beta}^{a} (0) T_{b}^{\nu} (0))
\int_{0}^{\sigma^{+}} k_{+}^{\beta} D \sigma^{'+} \label{equation3.68}\\
&+ (f_{c b}^{a} T_{a}^{\alpha} (0) - \tilde{f}_{\beta b}^{\alpha}
T_{c}^{\beta} (0) + \tilde{f}_{\beta c}^{\alpha} T_{b}^{\beta} (0))
\int_{0}^{\sigma^{+}} A_{+}^{c} D^{'} \sigma^{'+} + H.O. \notag
\end{align}
where all initial values are chosen to be identity. Therefore
pseudodual nonlocal currents on $\tilde{\textbf{m}}$  can be written
as
\begin{align} \label{equation3.69}
\tilde{k}_{+}^{\alpha} = &k_{+}^{\alpha} +
T_{b}^{\alpha}(0)A_{+}^{b} + (f_{\beta \lambda}^{a} T_{a}^{\alpha}
(0) - \tilde{f}_{a \lambda}^{\alpha} T_{\beta}^{a} (0) +
\tilde{f}_{a \beta}^{\alpha} T_{\lambda}^{a} (0)) k_{+}^{\lambda}
\int_{0}^{\sigma^{+}} k_{+}^{\beta} D \sigma^{'+} \notag\\ &+ (f_{b
\beta}^{\alpha} - \tilde{f}_{b \beta}^{\alpha} + \tilde{f}_{a
\nu}^{\alpha} T_{b}^{\nu} (0) T_{\beta}^{a} (0))
\int_{0}^{\sigma^{+}} (A_{+}^{b} (\sigma^{'+})
k_{+}^{\beta}(\sigma^{+}) - k_{+}^{\beta} (\sigma^{'+}) A_{+}^{b}
(\sigma^{+})) d\sigma^{'+} \notag\\ &+ (f_{c b}^{a} T_{a}^{\alpha}
(0) - \tilde{f}_{\beta b}^{\alpha} T_{c}^{\beta} (0) +
\tilde{f}_{\beta c}^{\alpha} T_{b}^{\beta} (0)) A_{+}^{b}
\int_{0}^{\sigma^{+}} A_{+}^{c} D^{'} \sigma^{'+} + H.O.
\end{align}
\begin{align} \label{equation3.70}
\tilde{k}_{-}^{\alpha} = &- k_{-}^{\alpha} -
T_{b}^{\alpha}(0)A_{-}^{b} - (f_{\beta \lambda}^{a} T_{a}^{\alpha} -
\tilde{f}_{a \lambda}^{\alpha} T_{\beta}^{a} (0) + \tilde{f}_{a
\beta}^{\alpha} T_{\lambda}^{a} (0)) k_{-}^{\lambda}
\int_{0}^{\sigma^{+}} k_{+}^{\beta} D \sigma^{'+} \notag\\ &+
(f_{\beta b}^{\alpha} + \tilde{f}_{b \beta}^{\alpha} - \tilde{f}_{a
\nu}^{\alpha} T_{b}^{\nu} (0) T_{\beta}^{a} (0))
\int_{0}^{\sigma^{+}} (A_{+}^{b} (\sigma^{'+}) k_{-}^{\beta}
(\sigma^{+}) - k_{+}^{\beta} (\sigma^{'+}) A_{-}^{b} (\sigma^{+})) d
\sigma^{'+} \notag\\ &- (f_{c b}^{a} T_{a}^{\alpha} (0) -
\tilde{f}_{\beta b}^{\alpha} T_{c}^{\beta} (0) + \tilde{f}_{\beta
c}^{\alpha} T_{b}^{\beta} (0)) A_{-}^{b} \int_{0}^{\sigma^{+}}
A_{+}^{c} D^{'} \sigma^{'+} + H.O.
\end{align}
Conservation laws of these currents up to the second order terms are
obvious. Now we consider pseudoduality equations on
$\textbf{h}$-space (\ref{equation3.44}). We take $\partial_{-}$ of
first equation, and we obtain
\begin{equation} \label{equation3.71}
(\partial_{-}T_{b}^{a})A_{+}^{b} +
(\partial_{-}T_{\alpha}^{a})k_{+}^{\alpha} = 0
\end{equation}
Hence we get $\partial_{-}T_{b}^{a} = \partial_{-}T_{\alpha}^{a} =
0$, which implies that $T_{b}^{a}$ and $T_{\alpha}^{a}$ don't depend
on $\sigma^{-}$. Taking $\partial_{+}$ of second equation we get the
following equation
\begin{align}
[\tilde{A}_{-}, \tilde{A}_{+}]^{a} + [\tilde{k}_{-},
\tilde{k}_{+}]^{a} = &- (\partial_{+} T_{b}^{a}) A_{-}^{b} -
T_{b}^{a} [A_{-}, A_{+}]^{b} - T_{b}^{a}
[k_{-}, k_{+}]^{b} \notag\\
&- (\partial_{+} T_{\alpha}^{a})k_{-}^{\alpha} - T_{\alpha}^{a}
[k_{-}, A_{+}]^{\alpha} - T_{\alpha}^{a} [A_{-}, k_{+}]^{\alpha}
\label{equation3.72}
\end{align}
We replace $\tilde{A}_{-}$ and $\tilde{k}_{-}$ in this equation to
obtain the following results
\begin{align}
\partial_{+} T_{d}^{a} = (T_{b}^{a} f_{e d}^{b} - \tilde{f}_{b c}^{a} T_{e}^{b} T_{d}^{c} - \tilde{f}_{\alpha \beta}^{a} T_{e}^{\alpha}
T_{d}^{\beta}) A_{+}^{e} + (T_{\alpha}^{a} f_{\lambda d}^{\alpha} -
\tilde{f}_{b c}^{a} T_{\lambda}^{b} T_{d}^{c} - \tilde{f}_{\alpha
\beta}^{a} T_{\lambda}^{\alpha} T_{d}^{\beta}) k_{+}^{\lambda}
\label{equation3.73}\\
\partial_{+} T_{\nu}^{a} = (T_{b}^{a} f_{\lambda \nu}^{b} - \tilde{f}_{b c}^{a} T_{\lambda}^{b} T_{\nu}^{c} - \tilde{f}_{\alpha \beta}^{a} T_{\lambda}^{\alpha}
T_{\nu}^{\beta}) k_{+}^{\lambda} + (T_{\alpha}^{a} f_{d
\nu}^{\alpha} - \tilde{f}_{b c}^{a} T_{d}^{b} T_{\nu}^{c} -
\tilde{f}_{\alpha \beta}^{a} T_{d}^{\alpha} T_{\nu}^{\beta})
A_{+}^{d} \label{equation3.74}
\end{align}
We again want to find solutions up to the second order terms, so we
only use initial values to get
\begin{align}
T_{d}^{a} (\sigma^{+}) = &T_{d}^{a} (0) + (f_{e d}^{a} -
\tilde{f}_{e d}^{a} - \tilde{f}_{\alpha \beta}^{a} T_{e}^{\alpha}
(0) T_{d}^{\beta} (0)) \int_{0}^{\sigma^{+}} A_{+}^{e} D^{'} \sigma^{'+} \label{equation3.75}\\
&+ (T_{\alpha}^{a} (0) f_{\lambda d}^{\alpha} - \tilde{f}_{b d}^{a}
T_{\lambda}^{b} (0) - \tilde{f}_{\lambda \beta}^{a} T_{d}^{\beta}
(0)) \int_{0}^{\sigma^{+}} k_{+}^{\lambda} D \sigma^{'+} + H.O.
\notag
\end{align}
\begin{align}
T_{\nu}^{a} (\sigma^{+}) = &T_{\nu}^{a} (0) + (f_{\lambda \nu}^{a} -
\tilde{f}_{\lambda \nu}^{a} - \tilde{f}_{b c}^{a} T_{\lambda}^{b}
(0) T_{\nu}^{c} (0)) \int_{0}^{\sigma^{+}} k_{+}^{\lambda} D \sigma^{'+} \label{equation3.76}\\
&+ (T_{\alpha}^{a} (0) f_{d \nu}^{\alpha} - \tilde{f}_{d c}^{a}
T_{\nu}^{c} (0) - \tilde{f}_{\alpha \nu}^{a} T_{d}^{\alpha} (0))
\int_{0}^{\sigma^{+}} A_{+}^{d} D^{'} \sigma^{'+} + H.O. \notag
\end{align}
Thus pseudodual fields up to the second order terms on $H$ space
will be
\begin{align}
\tilde{A}_{+}^{a} &= A_{+}^{a} + T_{\lambda}^{a} (0) k_{+}^{\lambda}
+ (f_{e d}^{a} - \tilde{f}_{e d}^{a} - \tilde{f}_{\alpha \beta}^{a}
T_{e}^{\alpha} (0) T_{d}^{\beta} (0)) A_{+}^{d}
\int_{0}^{\sigma^{+}} A_{+}^{e} D^{'} \sigma^{'+} \notag\\
&+ (T_{\alpha}^{a} (0) f_{\lambda d}^{\alpha} - \tilde{f}_{b d}^{a}
T_{\lambda}^{b} (0) - \tilde{f}_{\lambda \beta}^{a} T_{d}^{\beta}
(0)) \int_{0}^{\sigma^{+}} (k_{+}^{\lambda} (\sigma^{'+}) A_{+}^{d}
(\sigma^{+}) - A_{+}^{d} (\sigma^{'+}) k_{+}^{\lambda} (\sigma^{+}))
d \sigma^{'+} \notag\\
&+ (f_{\lambda \nu}^{a} - \tilde{f}_{\lambda \nu}^{a} - \tilde{f}_{b
c}^{a} T_{\lambda}^{b} (0) T_{\nu}^{c} (0)) k_{+}^{\nu}
\int_{0}^{\sigma^{+}} k_{+}^{\lambda} D \sigma^{'+} + H.O.
\label{equation3.77}
\end{align}
\begin{align}
\tilde{A}_{-}^{a} &= - A_{-}^{a} - T_{\lambda}^{a} (0)
k_{-}^{\lambda} - (f_{e d}^{a} - \tilde{f}_{e d}^{a} -
\tilde{f}_{\alpha \beta}^{a} T_{e}^{\alpha} (0) T_{d}^{\beta} (0))
A_{-}^{d} \int_{0}^{\sigma^{+}} A_{+}^{e} D^{'} \sigma^{'+} \notag\\
&- (T_{\alpha}^{a} (0) f_{\lambda d}^{\alpha} - \tilde{f}_{b d}^{a}
T_{\lambda}^{b} (0) - \tilde{f}_{\lambda \beta}^{a} T_{d}^{\beta}
(0)) \int_{0}^{\sigma^{+}} (k_{+}^{\lambda} (\sigma^{'+}) A_{-}^{d}
(\sigma^{+}) - A_{+}^{d} (\sigma^{'+}) k_{-}^{\lambda} (\sigma^{+}))
d \sigma^{'+} \notag\\
&- (f_{\lambda \nu}^{a} - \tilde{f}_{\lambda \nu}^{a} - \tilde{f}_{b
c}^{a} T_{\lambda}^{b} (0) T_{\nu}^{c} (0)) k_{-}^{\nu}
\int_{0}^{\sigma^{+}} k_{+}^{\lambda} D \sigma^{'+} + H.O.
\label{equation3.78}
\end{align}
It is obvious that conservation laws (\ref{equation3.42}) and
(\ref{equation3.43}) up to the second order terms are satisfied
\begin{align}
\tilde{D}_{-}^{'}\tilde{A}_{+}^{a} = &0 \label{equation3.79}\\
\tilde{D}_{+}^{'}\tilde{A}_{-}^{a} = &- [A_{-},
A_{+}]_{\tilde{G}}^{a} - [k_{-}, k_{+}]_{\tilde{G}}^{a} - [T
(0)A_{-}, T(0)A_{+}]_{\tilde{G}}^{a} - [A_{-},
T(0)k_{+}]_{\tilde{G}}^{a} \notag\\ &- [T(0)k_{-},
A_{+}]_{\tilde{G}}^{a} - [T(0)A_{-}, k_{+}]_{\tilde{G}}^{a} -
[k_{-}, T(0)A_{+}]_{\tilde{G}}^{a} \notag\\ &- [T(0)k_{-},
T(0)k_{+}]_{\tilde{G}}^{a} + H.O. \label{equation3.80}
\end{align}

\subsection{Dual Symmetric Spaces and Further Constraints}
\label{sec:DSSFC}

It is well-known \cite{alvarez1, oneill} that two normal symmetric
spaces are dual symmetric spaces if there exist

1. a Lie algebra isomorphism $S$ : $\textbf{h} \longrightarrow
\tilde{\textbf{h}}$ such that $\tilde{Q}(SV, SW) = - Q (V, W)$ for
all $V, W \in \textbf{h}$, and $Q$ is inner product.

2. a linear isometry $T : \textbf{m} \longrightarrow
\tilde{\textbf{m}}$ such that $[TX, TY] = -S[X, Y]$ for all $X, Y
\in \textbf{m}$.

Item (1) tells us that brackets in $\textbf{h}$ and
$\tilde{\textbf{h}}$ are the same while item (2) tells us that inner
products in $\textbf{m}$ and $\tilde{\textbf{m}}$ are the same. Item
(1) yields the result $f_{c b}^{a} = \tilde{f}_{c b}^{a}$ for
non-mixing pseudoduality, which leads $T_{b}^{a}$ to be a constant.
Hence pseudoduality transformations will simply be
\begin{align}
\tilde{k}_{\pm}^{\alpha} = \pm k_{\pm}^{\alpha} \label{equation3.81}\\
\tilde{A}_{\pm}^{a} = \pm A_{\pm}^{a} \label{equation3.82}
\end{align}
with the bracket relations (\ref{equation3.57})-(\ref{equation3.60})
given by
\begin{align}
[\tilde{k}_{+}, \tilde{A}_{-}]^{\alpha} &= - [k_{+}, A_{-}]^{\alpha} \label{equation3.83}\\
[\tilde{k}_{-}, \tilde{A}_{+}]^{\alpha} &= - [k_{-}, A_{+}]^{\alpha} \label{equation3.84}\\
[\tilde{k}_{+}, \tilde{k}_{-}]^{a} &= - [k_{+}, k_{-}]^{a} \label{equation3.85}\\
[\tilde{A}_{+}, \tilde{A}_{-}]^{a} &= - [A_{+}, A_{-}]^{a}
\label{equation3.86}
\end{align}

On the other hand one can write the following bracket relations
between pseudodual target spaces for the mixing pseudoduality case
\begin{align}
[\tilde{k}_{-}, \tilde{A}_{+}]^{\alpha} + [\tilde{A}_{-},
\tilde{k}_{+}]^{\alpha} = &- T_{\beta}^{\alpha} [k_{-},
A_{+}]^{\beta} - T_{\beta}^{\alpha} [A_{-}, k_{+}]^{\beta}
\label{equation3.87}\\ &- T_{a}^{\alpha} [A_{-}, A_{+}]^{a} -
T_{a}^{\alpha} [k_{-}, k_{+}]^{a} \notag\\
[\tilde{A}_{-}, \tilde{A}_{+}]^{a} + [\tilde{k}_{-},
\tilde{k}_{+}]^{a} = &- T_{b}^{a} [A_{-}, A_{+}]^{b} - T_{b}^{a}
[k_{-}, k_{+}]^{b} \label{equation3.88}\\
&- T_{\alpha}^{a} [k_{-}, A_{+}]^{\alpha} - T_{\alpha}^{a} [A_{-},
k_{+}]^{\alpha} \notag
\end{align}
which in turn leads to relations of connection two-forms between
symmetric and corresponding H-spaces, which is consistent with the
result found in section 5 (\ref{sec5:PSSMSS}). These equations
produce that all components of the pseudoduality map $T$ must be
constant, and we choose them to be identity. Hence pseudoduality
equations will simply be
\begin{align}
\tilde{k}_{\pm}^{\alpha} = \pm k_{\pm}^{\alpha} \pm T_{a}^{\alpha}
(0)
A_{\pm}^{a} \label{equation3.89}\\
\tilde{A}_{\pm}^{a} = \pm A_{\pm}^{a} \pm T_{\alpha}^{a} (0)
k_{\pm}^{\alpha} \label{equation3.90}
\end{align}

\subsection{An Example} \label{sec3:example}

We consider the Lie groups we used in the previous section. We saw
that invariant subspace of $SO(n + 1)$ is $1 \times SO(n)$. We pick
$H$ space as $SO(n)$. Hence our symmetric space is $M = \frac{SO(n +
1)}{SO(n)}$. The Lie algebra $\textbf{g }= so(n + 1)$ can be written
as
\begin{align} \label{equation3.91}
so(n + 1) = \left(
              \begin{array}{cc}
                a & b \\
                -b^{t} & c \\
              \end{array}
            \right) \ \ \ \ \ \ \begin{array}{c}
                                  a = 1 \times 1 \\
                                  b = 1 \times n \\
                                  c = n \times n
                                \end{array}
\end{align}
which can be split as
\begin{equation} \label{equation3.92}
\left(
  \begin{array}{cc}
    a & b \\
    -b^{t} & c \\
  \end{array}
\right) = \left(
            \begin{array}{cc}
              a & 0 \\
              0 & c \\
            \end{array}
          \right) + \left(
                      \begin{array}{cc}
                        0 & b \\
                        -b^{t} & 0 \\
                      \end{array}
                    \right) \ \ \ \  \ \ \ \ \ \textbf{g} = \textbf{h} \oplus \textbf{m}
\end{equation}

Let $Y \in \textbf{g}$, $X \in \textbf{h}$, and $Z \in \textbf{m}$.
Then, $D^{'}Z = 0$ and $DX = 0$. Using the expansions
(\ref{equation3.14}) and (\ref{equation3.15}), we may write the
following expressions
\begin{align}
k_{+}^{\alpha} &= D_{+}Z_{L}^{\alpha} - \frac{1}{2} [X_{L}, D_{+} Z_{L}]^{\alpha} - \frac{1}{2} [Z_{L}, D_{+}^{'} X_{L}]^{\alpha} + H.O. \label{equation3.93}\\
A_{+}^{a} &= D_{+}^{'}X_{L}^{a} - \frac{1}{2} [X_{L}, D_{+}^{'}
X_{L}]^{a} - \frac{1}{2} [Z_{L}, D_{+} Z_{L}]^{a} + H.O.
\label{equation3.94}\\
k_{-}^{\alpha} &= D_{-} Z_{R}^{\alpha} - [X_{L}, D_{-}
Z_{R}]^{\alpha} - [Z_{L}, D_{-}^{'} X_{R}]^{\alpha} - \frac{1}{2}
[X_{R}, D_{-} Z_{R}]^{\alpha} \label{equation3.95}\\ &- \frac{1}{2}
[Z_{R}, D_{-}^{'} X_{R}]^{\alpha} + H.O. \notag\\
A_{-}^{a} &= D_{-}^{'} X_{R}^{a} - [X_{L}, D_{-}^{'}X_{R}]^{a} -
[Z_{L}, D_{-} Z_{R}]^{a} - \frac{1}{2} [X_{R}, D_{-}^{'} X_{R}]^{a}
\label{equation3.96}\\ &- \frac{1}{2} [Z_{R}, D_{-} Z_{R}]^{a} +
H.O. \notag
\end{align}
We describe solutions $X = \sum_{n = 1}^{\infty} \varepsilon^{n}
x_{n}$ and $Z = \sum_{n = 1}^{\infty} \varepsilon^{n} z_{n}$, where
$\varepsilon$ is a small parameter. It is clear that equations of
motion (\ref{equation3.40})-(\ref{equation3.43}) for all orders of
$\varepsilon$ are satisfied. In the following calculations we are
going to use expressions up to the order of $\varepsilon^{2}$ for
simplicity.

Now we consider dual symmetric space $\tilde{M} = \frac{SO(n,
1)}{SO(n)}$, where $\tilde{H} = SO(n)$. Lie algebra
$\tilde{\textbf{g}} = so(n, 1)$ is written as
\begin{align} \label{equation3.97}
so(n, 1) = \left(
              \begin{array}{cc}
                \tilde{a} & \tilde{b} \\
                \tilde{b}^{t} & \tilde{c} \\
              \end{array}
            \right) \ \ \ \ \ \ \begin{array}{c}
                                  \tilde{a} = 1 \times 1 \\
                                  \tilde{b} = 1 \times n \\
                                  \tilde{c} = n \times n
                                \end{array}
\end{align}
which is split as
\begin{equation} \label{equation3.98}
\left(
  \begin{array}{cc}
    \tilde{a} & \tilde{b} \\
    \tilde{b}^{t} & \tilde{c} \\
  \end{array}
\right) = \left(
            \begin{array}{cc}
              \tilde{a} & 0 \\
              0 & \tilde{c} \\
            \end{array}
          \right) + \left(
                      \begin{array}{cc}
                        0 & \tilde{b} \\
                        \tilde{b}^{t} & 0 \\
                      \end{array}
                    \right) \ \ \ \  \ \ \ \ \ \tilde{\textbf{g}} = \tilde{\textbf{h}} \oplus \tilde{\textbf{m}}
\end{equation}

Let $\tilde{Y} = \tilde{X} + \tilde{Z}$, where $\tilde{Y} \in
\tilde{\textbf{g}}$, $\tilde{X} \in \tilde{\textbf{h}}$, and
$\tilde{Z} \in \tilde{\textbf{m}}$. We get the same fields as
equations (\ref{equation3.93})-(\ref{equation3.96}) with tilde.
Equations of motion will be the same with tilde. We may now find
pseudodual fields using our expressions found above. We note that
because of the special form of our Lie groups, mixing components of
the map $T$ vanishes, and we simply get non-mixing pseudoduality
condition.

We insert our expressions into equations (\ref{equation3.81}) and
(\ref{equation3.82}) to get infinitely many pseudoduality relations.
Up to the order of $\varepsilon^{2}$ terms equation
(\ref{equation3.81}) will be
\begin{align}
\tilde{D}_{+} \tilde{z}_{L1}^{\alpha} = D_{+} z_{L1}^{\alpha} \ \ \
\ \ \ \ \ \ \ \tilde{D}_{-} \tilde{z}_{R1}^{\alpha} = - D_{-}
z_{R1}^{\alpha} \label{equation3.99}
\end{align}
\begin{align}
\tilde{D}_{+} \tilde{z}_{L2}^{\alpha} - \frac{1}{2} [\tilde{x}_{L1},
\tilde{D}_{+}\tilde{z}_{L1}]^{\alpha} - \frac{1}{2} [\tilde{z}_{L1},
\tilde{D}_{+}^{'} \tilde{x}_{L1}]^{\alpha} = D_{+} z_{L2}^{\alpha} -
\frac{1}{2} [x_{L1}, D_{+}z_{L1}]^{\alpha} - \frac{1}{2} [z_{L1},
D_{+}^{'} x_{L1}]^{\alpha} \notag
\end{align}
\begin{align}
&\tilde{D}_{-} \tilde{z}_{R2}^{\alpha} - [\tilde{x}_{L1},
\tilde{D}_{-} \tilde{z}_{R1}]^{\alpha} - [\tilde{z}_{L1},
\tilde{D}_{-}^{'} \tilde{x}_{R1}]^{\alpha} - \frac{1}{2}
[\tilde{x}_{R1}, \tilde{D}_{-}\tilde{z}_{R1}]^{\alpha} - \frac{1}{2}
[\tilde{z}_{R1}, \tilde{D}_{-}^{'} \tilde{x}_{R1}]^{\alpha} =
\notag\\ &- D_{-} z_{R2}^{\alpha} + [x_{L1}, D_{-} z_{R1}]^{\alpha}
+ [z_{L1}, D_{-}^{'} x_{R1}]^{\alpha} + \frac{1}{2} [x_{R1},
D_{-}z_{R1}]^{\alpha} + \frac{1}{2} [z_{R1}, D_{-}^{'}
x_{R1}]^{\alpha} \notag
\end{align}
and equation (\ref{equation3.82}) will be
\begin{align}
\tilde{D}_{+}^{'} \tilde{x}_{L1}^{a} = D_{+}^{'} x_{L1}^{a} \ \ \ \
\ \ \ \ \ \ \tilde{D}_{-}^{'} \tilde{x}_{R1}^{a} = - D_{-}^{'}
x_{R1}^{a} \label{equation3.100}
\end{align}
\begin{align}
\tilde{D}_{+}^{'} \tilde{x}_{L2}^{a} - \frac{1}{2} [\tilde{x}_{L1},
\tilde{D}_{+}^{'} \tilde{x}_{L1}]^{a} - \frac{1}{2} [\tilde{z}_{L1},
\tilde{D}_{+} \tilde{z}_{L1}]^{a} = D_{+}^{'} x_{L2}^{a} -
\frac{1}{2} [x_{L1}, D_{+}^{'} x_{L1}]^{a} - \frac{1}{2} [z_{L1},
D_{+} z_{L1}]^{a} \notag
\end{align}
\begin{align}
&\tilde{D}_{-}^{'} \tilde{x}_{R2}^{a} - [\tilde{x}_{L1},
\tilde{D}_{-}^{'} \tilde{x}_{R1}]^{a} - [\tilde{z}_{L1},
\tilde{D}_{-} \tilde{z}_{R1}]^{a} - \frac{1}{2} [\tilde{x}_{R1},
\tilde{D}_{-}^{'} \tilde{x}_{R1}]^{a} -\frac{1}{2} [\tilde{z}_{R1},
\tilde{D}_{-} \tilde{z}_{R1}]^{a} = \notag\\ &- D_{-}^{'} x_{R2}^{a}
+ [x_{L1}, D_{-}^{'} x_{R1}]^{a} + [z_{L1}, D_{-} z_{R1}]^{a} +
\frac{1}{2} [x_{R1}, D_{-}^{'} x_{R1}]^{a} + \frac{1}{2} [z_{R1},
D_{-} z_{R1}]^{a} \notag
\end{align}

Since we know
\begin{equation}
D_{\pm}z_{n} = \left(
           \begin{array}{cc}
             0 & D_{\pm}b_{n} \\
             -D_{\pm}b_{n}^{t} & 0 \\
           \end{array}
         \right) \ \ \ \ \ D_{\pm}^{'}x_{n} = \left(
                                          \begin{array}{cc}
                                            D_{\pm}^{'}a_{n} & 0 \\
                                            0 & D_{\pm}^{'}c_{n} \\
                                          \end{array}
                                        \right) \notag
\end{equation}
\begin{equation}
[x_{1}, D_{\pm}^{'}x_{1}] = \left(
                    \begin{array}{cc}
                      [a_{1}, D_{\pm}^{'}a_{1}] & 0 \\
                      0 & [c_{1}, D_{\pm}^{'}c_{1}] \\
                    \end{array}
                  \right) \notag
\end{equation}
\begin{equation}
[z_{1}, D_{\pm} z_{1}] = \left(
                    \begin{array}{cc}
                      (D_{\pm} b_{1}) b_{1}^{t} - b_{1} (D_{\pm} b_{1}^{t}) & 0 \\
                      0 & (D_{\pm} b_{1}^{t}) b_{1} - b_{1}^{t} (D_{\pm} b_{1}) \\
                    \end{array}
                  \right) \notag
\end{equation}
\begin{equation}
[x_{1}, D_{\pm}z_{1}] = \left(
                            \begin{array}{cc}
                               0 & a_{1}D_{\pm}b_{1} - (D_{\pm}b_{1})c_{1} \\
                               -c_{1}(D_{\pm}b_{1}^{t}) + (D_{\pm}b_{1}^{t})a_{1} & 0 \\
                            \end{array}
                      \right) \notag
\end{equation}
\begin{equation}
[z_{1}, D_{\pm}^{'} x_{1}] = \left(
                            \begin{array}{cc}
                               0 & b_{1}D_{\pm}^{'} c_{1} - (D_{\pm}^{'} a_{1})b_{1} \\
                               -b_{1}^{t}(D_{\pm}^{'} a_{1}) + (D_{\pm}^{'} c_{1})b_{1}^{t} & 0 \\
                            \end{array}
                      \right) \notag
\end{equation}
One can write similar expressions on the pseudodual space replacing
each term with tilded terms. Only exception is that we switch
$b_{n}^{t}$ with $- \tilde{b}_{n}^{t}$ so that we get the convenient
lie algebra on tilded space. Therefore pseudoduality equations above
(\ref{equation3.99}) and (\ref{equation3.100}) will give the
following expressions
\begin{equation}
\tilde{D}_{+} \tilde{b}_{L1} = D_{+} b_{L1} \ \ \ \ \ \tilde{D}_{+}
\tilde{b}_{L1}^{t} = - D_{+} b_{L1}^{t} \notag
\end{equation}
\begin{equation}
\tilde{D}_{-} \tilde{b}_{R1} = - D_{-} b_{R1} \ \ \ \ \
\tilde{D}_{-} \tilde{b}_{R1}^{t} = D_{-} b_{R1}^{t} \notag
\end{equation}
\begin{equation}
\tilde{D}_{+}^{'} \tilde{a}_{L1} = D_{+}^{'} a_{L1} \ \ \ \ \
\tilde{D}_{+}^{'} \tilde{c}_{L1} = D_{+}^{'} c_{L1} \notag
\end{equation}
\begin{equation}
\tilde{D}_{-}^{'} \tilde{a}_{R1} = - D_{-}^{'} a_{R1} \ \ \ \ \
\tilde{D}_{-}^{'} \tilde{c}_{R1} = - D_{-}^{'} c_{R1} \notag
\end{equation}
\begin{align}
\tilde{D}_{+} \tilde{b}_{L2} = D_{+} b_{L2} &+ \frac{1}{2}
\{(\tilde{a}_{L1} - a_{L1}) D_{+} b_{L1} - D_{+} b_{L1}
(\tilde{c}_{L1} - c_{L1})\} \notag\\ &+ \frac{1}{2}
\{(\tilde{b}_{L1} - b_{L1}) D_{+}^{'} c_{L1} - D_{+}^{'} a_{L1}
(\tilde{b}_{L1} - b_{L1})\} \notag
\end{align}
\begin{align}
\tilde{D}_{+} \tilde{b}_{L2}^{t} = - D_{+} b_{L2}^{t} &- \frac{1}{2}
\{(\tilde{c}_{L1} - c_{L1}) D_{+} b_{L1}^{t} - D_{+} b_{L1}^{t}
(\tilde{a}_{L1} - a_{L1})\} \notag\\ &+ \frac{1}{2}
\{(\tilde{b}_{L1}^{t} + b_{L1}^{t}) D_{+}^{'} a_{L1} - D_{+}^{'}
c_{L1} (\tilde{b}_{L1}^{t} + b_{L1}^{t})\} \notag
\end{align}
\begin{align}
\tilde{D}_{+}^{'} \tilde{a}_{L2} = D_{+}^{'} a_{L2} &+ \frac{1}{2}
[(\tilde{a}_{L1} - a_{L1}), D_{+}^{'} a_{L1}] \notag\\ &-
\frac{1}{2} \{D_{+} b_{L1} (b_{L1}^{t} + \tilde{b}_{L1}^{t}) -
(b_{L1} - \tilde{b}_{L1}) D_{+} b_{L1}^{t}\} \notag
\end{align}
\begin{align}
\tilde{D}_{+}^{'} \tilde{c}_{L2} = D_{+}^{'} c_{L2} &+ \frac{1}{2}
[(\tilde{c}_{L1} - c_{L1}), D_{+}^{'} c_{L1}] \notag\\ &-
\frac{1}{2} \{D_{+} b_{L1}^{t} (b_{L1} - \tilde{b}_{L1}) -
(b_{L1}^{t} + \tilde{b}_{L1}^{t}) D_{+} b_{L1}\} \notag
\end{align}
\begin{align}
\tilde{D}_{-} \tilde{b}_{R2} = - D_{-} b_{R2} &+ \{(a_{L1} -
\tilde{a}_{L1}) + \frac{1}{2} (a_{R1} - \tilde{a}_{R1}) \} D_{-}
b_{R1} \notag\\ &- D_{-}b_{R1}\{(c_{L1} - \tilde{c}_{L1}) +
\frac{1}{2} (c_{R1} - \tilde{c}_{R1})\} \notag\\ &+ \{(b_{L1} -
\tilde{b}_{L1}) + \frac{1}{2} (b_{R1} - \tilde{b}_{R1})\} D_{-}^{'}
c_{R1} \notag\\ &- D_{-}^{'} a_{R1} \{(b_{L1} - \tilde{b}_{L1}) +
\frac{1}{2} (b_{R1} - \tilde{b}_{R1})\} \notag
\end{align}
\begin{align}
\tilde{D}_{-} \tilde{b}_{R2}^{t} = D_{-} b_{R2}^{t} &- \{(c_{L1} -
\tilde{c}_{L1}) + \frac{1}{2} (c_{R1} - \tilde{c}_{R1}) \} D_{-}
b_{R1}^{t} \notag\\ &+ D_{-}b_{R1}^{t} \{(a_{L1} - \tilde{a}_{L1}) +
\frac{1}{2} (a_{R1} - \tilde{a}_{R1})\} \notag\\ &- \{(b_{L1}^{t} +
\tilde{b}_{L1}^{t}) + \frac{1}{2} (b_{R1}^{t} +
\tilde{b}_{R1}^{t})\} D_{-}^{'} a_{R1} \notag\\ &+ D_{-}^{'} c_{R1}
\{(b_{L1}^{t} + \tilde{b}_{L1}^{t}) + \frac{1}{2} (b_{R1}^{t} +
\tilde{b}_{R1}^{t})\} \notag
\end{align}
\begin{align}
\tilde{D}_{-}^{'} \tilde{a}_{R2} = - D_{-}^{'} a_{R2} &+ [(a_{L1} -
\tilde{a}_{L1}) + \frac{1}{2} (a_{R1} - \tilde{a}_{R1}), D_{-}^{'}
a_{R1}] \notag\\ &+ D_{-} b_{R1} \{(b_{L1}^{t} + \tilde{b}_{L1}^{t})
+ \frac{1}{2} (b_{R1}^{t} + \tilde{b}_{R1}^{t})\} \notag\\ &-
\{(b_{L1} - \tilde{b}_{L1}) + \frac{1}{2} (b_{R1} -
\tilde{b}_{R1})\} D_{-} b_{R1}^{t} \notag
\end{align}
\begin{align}
\tilde{D}_{-}^{'} \tilde{c}_{R2} = - D_{-}^{'} c_{R2} &+ [(c_{L1} -
\tilde{c}_{L1}) + \frac{1}{2} (c_{R1} - \tilde{c}_{R1}), D_{-}^{'}
c_{R1}] \notag\\ &+ D_{-} b_{R1}^{t} \{(b_{L1} - \tilde{b}_{L1}) +
\frac{1}{2} (b_{R1} + \tilde{b}_{R1})\} \notag\\ &- \{(b_{L1}^{t} +
\tilde{b}_{L1}^{t}) + \frac{1}{2} (b_{R1}^{t} +
\tilde{b}_{R1}^{t})\} D_{-} b_{R1} \notag
\end{align}
where tilded terms on the right hand sides can be replaced by
solving corresponding equations. One can obtain the conserved
nonlocal currents using these terms.

\section{Curvatures} \label{sec3:cur}

\subsection{Case I: Curvatures on $\textbf{g}$ and
$\tilde{\textbf{g}}$}

Let us find the curvatures related to symmetric spaces, and see the
relations between dual symmetric parts. We first consider the case
where H = id. We may choose orthonormal frame $\{J\}$ on the
pullback bundle $g^{*}(TG)$, where $J$ stands for both $J^{(R)}$ and
$J^{(L)}$. These currents satisfy the Maurer-Cartan equation
\begin{equation} \label{equation3.101}
dJ^{i} + \frac{1}{2}f_{jk}^{i}J^{j} \wedge J^{k} = 0
\end{equation}
where $w^{i} = J^{i}$ and $w_{k}^{i} = \frac{1}{2}f_{jk}^{i}J^{j}$
is the antisymmetric riemannian connection. Curvature can be found
using torsion free Cartan structural equations
\begin{align}
dw^{i} + w_{j}^{i} \wedge w^{j} &= 0 \label{equation3.102}\\
dw_{j}^{i} + w_{k}^{i} \wedge w_{j}^{k} &=
\frac{1}{2}R_{jkl}^{i}w^{k} \wedge w^{l} \label{equation3.103}
\end{align}
Substituting $w^{i} = J^{i}$ and $w_{j}^{i} =
\frac{1}{2}f_{kj}^{i}J^{k}$ into first equation gives us the
Maurer-Cartan equation (\ref{equation3.101}). Curvature tensor
associated with $\textbf{g}$ can be found using second equation
(\ref{equation3.103}),
\begin{equation} \label{equation3.104}
R_{jmn}^{i} = - \frac{1}{2}(f_{km}^{i}f_{nj}^{k} +
f_{kj}^{i}f_{mn}^{k}) = \frac{1}{2}f_{kn}^{i}f_{jm}^{k}
\end{equation}
where we used jacobi identity in the last equation,
$f_{k[m}^{i}f_{nj]}^{k} = 0$. We may find similar relations for
pseudodual space with tilde (just put $\tilde{~}$ on each term). To
relate curvature tensor on pseudodual space with regular space, we
use nonlocal expressions (\ref{equation3.26})-(\ref{equation3.29}).
Since both currents yield the same result, we just use
(\ref{equation3.26}) and (\ref{equation3.27}) for the final
expression. We may write $\tilde{J}^{i}$ in nonlocal terms as
\begin{equation} \label{equation3.105}
\tilde{J}^{i} = \varepsilon dy_{1}^{i} + \varepsilon^{2}[dy_{2}^{i}
+ \frac{1}{2}f_{jk}^{i}y_{1}^{i} \wedge dy_{1}^{k} -
\tilde{f}_{jk}^{i}y_{1}^{j} \wedge dy_{1}^{k}] + H.O.
\end{equation}
Hence $\tilde{w}^{i} = \tilde{J}^{i}$, and $\tilde{w}_{k}^{i}$ can
be written as
\begin{align}
\tilde{w}_{k}^{i} &= \frac{1}{2}\tilde{f}_{jk}^{i}\tilde{J}^{j}
\notag\\ &= \frac{\varepsilon}{2}\tilde{f}_{jk}^{i}dy_{1}^{j} +
\frac{\varepsilon^{2}}{2}\tilde{f}_{jk}^{i}[dy_{2}^{j} +
\frac{1}{2}f_{mn}^{j}y_{1}^{m} \wedge dy_{1}^{n} -
\tilde{f}_{mn}^{j}y_{1}^{m} \wedge dy_{1}^{n}] + H.O.
\label{equation3.106}
\end{align}
We plug $\tilde{w}^{i}$ and $\tilde{w}_{k}^{i}$ into the second
Cartan structural equation on pseudodual space in the form
\begin{equation} \label{equation3.107}
d\tilde{w}_{j}^{i} + \tilde{w}_{k}^{i} \wedge \tilde{w}_{j}^{k} =
\frac{1}{2}\tilde{R}_{jkl}^{i}\tilde{w}^{k} \wedge \tilde{w}^{l}
\end{equation}
to obtain the curvature expression
\begin{equation} \label{equation3.108}
\tilde{R}_{jmn}^{i} = \frac{1}{2}\tilde{f}_{kj}^{i}f_{mn}^{k} -
\tilde{f}_{kj}^{i}\tilde{f}_{mn}^{k} +
\frac{1}{2}\tilde{f}_{mk}^{i}\tilde{f}_{nj}^{k}
\end{equation}

Since by definition $\tilde{R}_{jmn}^{i}$ (\ref{equation3.104})can
also be written as
\begin{equation} \label{equation3.109}
\tilde{R}_{jmn}^{i} =
\frac{1}{2}\tilde{f}_{kn}^{i}\tilde{f}_{jm}^{k}
\end{equation}
we get a relation between structure constants on spaces $\textbf{g}$
and $\tilde{\textbf{g}}$
\begin{equation} \label{equation3.110}
\frac{1}{2}\tilde{f}_{kj}^{i}f_{mn}^{k} =
\frac{1}{2}\tilde{f}_{kj}^{i}\tilde{f}_{mn}^{k}
\end{equation}
where we used the jacobi identity
$\tilde{f}_{k[n}^{i}\tilde{f}_{jm]}^{k} = 0$. Though we do not set
$f_{mn}^{k}$ equal to $\tilde{f}_{mn}^{k}$, we may treat them on
equal footing, and use one for another interchangeably in paired
terms. Hence $\tilde{R}_{jmn}^{i}$ (\ref{equation3.108}) can be
written in nonlocal structure constants as
\begin{equation} \label{equation3.111}
\tilde{R}_{jmn}^{i} = -\frac{1}{2}f_{kn}^{i}f_{jm}^{k} = -
R_{jmn}^{i}
\end{equation}
where we used $f_{k[j}^{i}f_{nm]}^{k} = 0$ after setting tilde terms
with nontilde terms. We note that we obtained pseudodual space
curvature as the negative regular space curvature. This shows that
spaces are dual symmetric spaces as we expressed above.

\subsection{Case II: Curvatures on Decomposed Spaces}

Let us decompose the current as $J = J^{\alpha}t_{\alpha} +
J^{a}t_{a}$, where we use indices $\alpha, \beta, \gamma, ...$ for
$\textbf{m}$ space and indices $a, b, c, ...$ for \textbf{h} space,
and $t_{\alpha}$ and $t_{a}$ are corresponding generators. We can
write the commutation relations as
\begin{equation} \label{equation3.112}
[t_{a}, t_{b}] = f_{ab}^{c}t_{c} \ \ \ \ \ \ \ [t_{a}, t_{\beta}] =
f_{a\beta}^{\alpha}t_{\alpha} \ \ \ \ \ \ \ [t_{\alpha}, t_{\beta}]
= f_{\alpha \beta}^{c}t_{c}
\end{equation}
Maurer-Cartan equation (\ref{equation3.101}) can be decomposed as
\begin{align}
dJ^{a} + \frac{1}{2}f_{bc}^{a}J^{b} \wedge J^{c} +
\frac{1}{2}f_{\alpha\beta}^{a}J^{\alpha} \wedge J^{\beta} = 0 \ \ \
\ \ \ \ on\ \textbf{h}-space \label{equation3.113}\\
dJ^{\alpha} + f_{\beta a}^{\alpha}J^{\beta} \wedge J^{a} = 0 \ \ \ \
\ \ \ \ on\ \textbf{m}-space \label{equation3.114}
\end{align}
We can also decompose Cartan structural equations. Decomposition of
first structural equation gives us
\begin{align}
dw^{a} + w_{b}^{a} \wedge w^{b} + w_{\alpha}^{a} \wedge w^{\alpha} =
0 \ \ \ \ \ \ \ on\ \textbf{h}-space \label{equation3.115}\\
dw^{\alpha} + w_{\beta}^{\alpha} \wedge w^{\beta} + w_{a}^{\alpha}
\wedge w^{a} = 0 \ \ \ \ \ \ \ on\ \textbf{m}-space
\label{equation3.116}
\end{align}
comparison of these equations with the Maurer-Cartan equations
(\ref{equation3.113})-(\ref{equation3.114}) gives us the following
connections
\begin{align}
w^{a} = J^{a} \ \ \ \ \ w_{c}^{a} = \frac{1}{2}f_{bc}^{a}J^{b} \ \ \
\ \ w_{\beta}^{a} = \frac{1}{2}f_{\alpha \beta}^{a}J^{\alpha}
\label{equation3.117}\\
w^{\alpha} = J^{\alpha} \ \ \ \ \ w_{\beta}^{\alpha} =
\frac{1}{2}f_{a\beta}^{\alpha}J^{a} \ \ \ \ \ w_{a}^{\alpha} =
\frac{1}{2}f_{\beta a}^{\alpha}J^{\beta} \label{equation3.118}
\end{align}
Decomposition of second Cartan structural equation leads to the
following equations
\begin{align}
dw_{b}^{a} + w_{c}^{a} \wedge w_{b}^{c} + w_{\lambda}^{a} \wedge
w_{b}^{\lambda} = &\frac{1}{2}R_{bcd}^{a}w^{c} \wedge w^{d} +
\frac{1}{2}R_{bc\lambda}^{a} w^{c} \wedge w^{\lambda} \label{equation3.119}\\
&+ \frac{1}{2}R_{b\lambda c}^{a} w^{\lambda}
\wedge w^{c} + \frac{1}{2}R_{b\lambda \mu}^{a}w^{\lambda} \wedge w^{\mu} \notag\\
dw_{\alpha}^{a} + w_{c}^{a} \wedge w_{\alpha}^{c} + w_{\lambda}^{a}
\wedge w_{\alpha}^{\lambda} = &\frac{1}{2}R_{\alpha bc}^{a}w^{b}
\wedge w^{c} + \frac{1}{2}R_{\alpha b\beta}^{a} w^{b} \wedge
w^{\beta} \label{equation3.120}\\ &+ \frac{1}{2}R_{\alpha\beta
b}^{a} w^{\beta} \wedge w^{b} + \frac{1}{2}R_{\alpha\lambda
\mu}^{a}w^{\lambda} \wedge w^{\mu} \notag\\
dw_{\beta}^{\alpha} + w_{\gamma}^{\alpha} \wedge w_{\beta}^{\gamma}
+ w_{a}^{\alpha} \wedge w_{\beta}^{a} = &\frac{1}{2}R_{\beta
ab}^{\alpha}w^{a} \wedge w^{b} + \frac{1}{2}R_{\beta
a\gamma}^{\alpha} w^{a} \wedge w^{\gamma} \label{equation3.121}\\ &+
\frac{1}{2}R_{\beta\gamma a}^{\alpha} w^{\gamma} \wedge w^{a} +
\frac{1}{2}R_{\beta\lambda \mu}^{\alpha}w^{\lambda} \wedge w^{\mu}
\notag\\
dw_{a}^{\alpha} + w_{\gamma}^{\alpha} \wedge w_{a}^{\gamma} +
w_{b}^{\alpha} \wedge w_{a}^{b} = &\frac{1}{2}R_{abc}^{\alpha}w^{b}
\wedge w^{c} + \frac{1}{2}R_{ab\lambda}^{\alpha} w^{b} \wedge
w^{\lambda} \label{equation3.122}\\ &+ \frac{1}{2}R_{a\lambda
b}^{\alpha} w^{\lambda} \wedge w^{b} + \frac{1}{2}R_{a\lambda
\mu}^{\alpha}w^{\lambda} \wedge w^{\mu} \notag
\end{align}
Inserting (\ref{equation3.117}) and (\ref{equation3.118}) into
(\ref{equation3.119}) gives the following curvature components
\begin{align}
R_{bde}^{a} &= \frac{1}{2}(f_{dc}^{a}f_{eb}^{c} -
f_{cb}^{a}f_{de}^{c}) = \frac{1}{2}f_{ce}^{a}f_{bd}^{c}
\label{equation3.123}\\
R_{b\alpha\beta}^{a} &= \frac{1}{2}(f_{\alpha\lambda}^{a}f_{\beta
b}^{\lambda} - f_{cb}^{a}f_{\alpha\beta}^{c}) =
\frac{1}{2}f_{\lambda\beta}^{a}f_{b\alpha}^{\lambda}
\label{equation3.124}\\
R_{bc\lambda}^{a} &= R_{b\lambda c}^{a} = 0 \label{equation3.125}
\end{align}
where we used the jacobi identity $f_{c[d}^{a}f_{be]}^{c} = 0$ in
(\ref{equation3.123}), and
$f_{\lambda\alpha}^{a}f_{b\beta}^{\lambda} +
f_{cb}^{a}f_{\beta\alpha}^{c} + f_{\lambda\beta}^{a}f_{\alpha
b}^{\lambda}$ in (\ref{equation3.124}). Likewise
(\ref{equation3.120}) gives the following curvature components
\begin{align}
R_{\alpha c\lambda}^{a} &=
\frac{1}{2}(f_{cd}^{a}f_{\lambda\alpha}^{d} -
f_{\beta\alpha}^{a}f_{c\lambda}^{\beta}) =
\frac{1}{2}f_{\beta\lambda}^{a}f_{\alpha c}^{\beta}
\label{equation3.126}\\
R_{\alpha\lambda c}^{a} &=
\frac{1}{2}(f_{\lambda\beta}^{a}f_{c\alpha}^{\beta} -
f_{\beta\alpha}^{a}f_{\lambda c}^{\beta}) =
\frac{1}{2}f_{bc}^{a}f_{\alpha\lambda}^{b} \label{equation3.127}\\
R_{\alpha bc}^{a} &= R_{\alpha\lambda\mu}^{a} = 0
\label{equation3.128}
\end{align}
where we used the jacobi identity $f_{dc}^{a}f_{\alpha\lambda}^{d} +
f_{\beta\alpha}^{a}f_{\lambda c}^{\beta} +
f_{\beta\lambda}^{a}f_{c\alpha}^{\beta} = 0$ in
(\ref{equation3.126}), and $f_{\beta\lambda}^{a}f_{\alpha c}^{\beta}
+ f_{\beta\alpha}^{a}f_{c\lambda}^{\beta} +
f_{bc}^{a}f_{\lambda\alpha}^{b} = 0$ in (\ref{equation3.127}).
Equation (\ref{equation3.121}) produces the following curvature
components
\begin{align}
R_{\beta bc}^{\alpha} &=
\frac{1}{2}(f_{b\gamma}^{\alpha}f_{c\beta}^{\gamma} -
f_{a\beta}^{\alpha}f_{bc}^{a}) = \frac{1}{2}f_{\gamma
c}^{\alpha}f_{\beta b}^{\gamma} \label{equation3.129}\\
R_{\beta\lambda\mu}^{\alpha} &= \frac{1}{2}(f_{\lambda
a}^{\alpha}f_{\mu\beta}^{a} - f_{a\beta}^{\alpha}f_{\lambda\mu}^{a})
= \frac{1}{2}f_{a\mu}^{\alpha}f_{\beta\lambda}^{a} \label{equation3.130}\\
R_{\beta a\gamma}^{\alpha} &= R_{\beta\gamma a}^{\alpha} = 0
\label{equation3.131}
\end{align}
where we used the jacobi identity $f_{\gamma b}^{\alpha}f_{\beta
c}^{\gamma} + f_{a\beta}^{\alpha}f_{cb}^{a} + f_{\gamma
c}^{\alpha}f_{b\beta}^{\gamma} = 0$ in (\ref{equation3.129}), and
$f_{a\lambda}^{\alpha}f_{\beta\mu}^{a} +
f_{a\beta}^{\alpha}f_{\mu\lambda}^{a} +
f_{a\mu}^{\alpha}f_{\lambda\beta}^{a} = 0$ in (\ref{equation3.130}).
Finally, equation (\ref{equation3.122}) gives the following
curvature components
\begin{align}
R_{ac\lambda}^{\alpha} &= \frac{1}{2}(f_{c\beta}^{\alpha}f_{\lambda
a}^{\beta} - f_{\beta a}^{\alpha}f_{c\lambda}^{\beta}) =
\frac{1}{2}f_{b\lambda}^{\alpha}f_{ac}^{b} \label{equation3.132}\\
R_{a\lambda c}^{\alpha} &= \frac{1}{2}(f_{\lambda
b}^{\alpha}f_{ca}^{b} - f_{\beta a}^{\alpha}f_{\lambda c}^{\beta})
= \frac{1}{2}f_{\beta c}^{\alpha}f_{a\lambda}^{\beta} \label{equation3.133}\\
R_{abc}^{\alpha} &= R_{a\lambda\mu}^{\alpha} = 0
\label{equation3.134}
\end{align}
where we used the jacobi identity $f_{\beta
c}^{\alpha}f_{a\lambda}^{\beta} + f_{\beta a}^{\alpha}f_{\lambda
c}^{\beta} + f_{b\lambda}^{\alpha}f_{ca}^{b} = 0$ in
(\ref{equation3.132}), and $f_{b\lambda}^{\alpha}f_{ac}^{b} +
f_{\beta a}^{\alpha}f_{c\lambda}^{\beta} + f_{\beta
c}^{\alpha}f_{\lambda a}^{\beta} = 0$ in (\ref{equation3.133}).
Obviously we can write similar equations with tilde.

We want to write down curvature relations between symmetric spaces
($\textbf{m}$ and $\tilde{\textbf{m}}$) and corresponding closed
spaces ($\textbf{h}$ and $\tilde{\textbf{h}}$) on $\textbf{g}$ and
$\tilde{\textbf{g}}$. To realize this objective we will use the
bracket relations derived from pseudoduality equations. In case of
non-mixing pseudoduality, we will make use of bracket relation
(\ref{equation3.83})-(\ref{equation3.86}). After eliminating $A_{-}$
and $k_{-}$ terms we obtain the following relations between
connection one forms
\begin{align}
\tilde{w}_{a}^{\alpha} = w_{a}^{\alpha} \ \ \ \ \ \ \ \ \ \
\tilde{w}_{\beta}^{\alpha} = w_{\beta}^{\alpha} \label{equation3.135}\\
\tilde{w}_{\beta}^{a} = w_{\beta}^{a} \ \ \ \ \ \ \ \ \ \
\tilde{w}_{b}^{a} = w_{b}^{a} \label{equation3.136}
\end{align}
where we used the definitions (\ref{equation3.117}) and
(\ref{equation3.118}) for the connection two forms. Taking exterior
derivative of these connections we obtain the result
\begin{equation}
\tilde{R}_{BCD}^{A} = - R_{BCD}^{A} \label{equation3.137}
\end{equation}
where $A$, $B$, $C$ and $D$ represent indices corresponding to $M$
or $H$-space elements depending on which equation is used. But
curvature expressions found above restrict all curvature components
to exist. Therefore we will only have curvatures whose all indices
belongs to one space ($\textbf{m}$ or $\textbf{h}$) or being shared
equally, otherwise they do not exist. On the other hand when we
consider mixing pseudoduality, we observe that curvature components
mix. From the connection two-forms we obtain the relations
\begin{align}
\tilde{w}_{\beta}^{\alpha} + \tilde{w}_{a}^{\alpha} T_{\beta}^{a}(0)
=
w_{\beta}^{\alpha} + T_{a}^{\alpha}(0) w_{\beta}^{a} \label{equation3.138}\\
\tilde{w}_{\beta}^{\alpha} T_{b}^{\beta}(0) + \tilde{w}_{b}^{\alpha}
=  T_{a}^{\alpha} (0) w_{b}^{a} + w_{b}^{\alpha} \label{equation3.139}\\
\tilde{w}_{b}^{a} + \tilde{w}_{\beta}^{a} T_{b}^{\beta}(0) =
w_{b}^{a} + T_{\beta}^{a}(0) w_{b}^{\beta} \label{equation3.140}\\
\tilde{w}_{b}^{a} T_{\beta}^{b}(0) + \tilde{w}_{\beta}^{a} =
T_{\gamma}^{a} (0) w_{\beta}^{\gamma} + w_{\beta}^{a}
\label{equation3.141}
\end{align}
It is clear that once mixing isometries disappear we have
(\ref{equation3.135}) and (\ref{equation3.136}). Therefore curvature
relations will be
\begin{align}
\hat{R}_{B \mu \nu}^{A} = - (\bar{\tilde{R}}_{B \mu \nu}^{A} +
\bar{\tilde{R}}_{B \mu c}^{A} T_{\nu}^{c}(0) + \bar{\tilde{R}}_{B c
\nu}^{A}T_{\mu}^{c}(0) + \bar{\tilde{R}}_{B c d}^{A} T_{\mu}^{c}(0)
T_{\nu}^{d}(0))
\label{equation3.142}\\
\hat{R}_{B \mu d}^{A} = - (\bar{\tilde{R}}_{B \mu d}^{A} +
\bar{\tilde{R}}_{B c d}^{A} T_{\mu}^{c}(0) + \bar{\tilde{R}}_{B \mu
\nu}^{A}T_{d}^{\nu}(0) + \bar{\tilde{R}}_{B c \nu}^{A}
T_{\mu}^{c}(0) T_{d}^{\nu}(0))
\label{equation3.143}\\
\hat{R}_{B c \nu}^{A} = - (\bar{\tilde{R}}_{B c \nu}^{A} +
\bar{\tilde{R}}_{B c d}^{A} T_{\nu}^{d}(0) + \bar{\tilde{R}}_{B \mu
\nu}^{A}T_{c}^{\mu}(0) + \bar{\tilde{R}}_{B \mu d}^{A}
T_{c}^{\mu}(0) T_{\nu}^{d}(0))
\label{equation3.144}\\
\hat{R}_{B c d}^{A} = - (\bar{\tilde{R}}_{B c d}^{A} +
\bar{\tilde{R}}_{B \mu d}^{A} T_{c}^{\mu}(0) + \bar{\tilde{R}}_{B c
\mu}^{A}T_{d}^{\mu}(0) + \bar{\tilde{R}}_{B \mu \nu}^{A}
T_{c}^{\mu}(0) T_{d}^{\nu}(0)) \label{equation3.145}
\end{align}
where we defined $\hat{R}_{\lambda \mu \nu}^{\alpha} \equiv
R_{\lambda \mu \nu}^{\alpha} + T_{a}^{\alpha}(0) R_{\lambda \mu
\nu}^{a}$ and $\bar{\tilde{R}}_{\lambda \mu \nu}^{\alpha} \equiv
\tilde{R}_{\lambda \mu \nu}^{\alpha} + \tilde{R}_{b \mu
\nu}^{\alpha} T_{\lambda}^{b}(0)$, and $A$, $B$ represent indices
for $\textbf{m}$ or $\textbf{h}$-spaces. Obviously if all mixing
parts are set to zero we obtain the simplest case
(\ref{equation3.137}).

\section{One Loop Renormalization Group
$\beta$-function}\label{sec3:ren}

It is noted that renormalization group $\beta$-function to one-loop
order \cite{ketov1} is given by
\begin{equation} \label{equation3.146}
\beta_{mn} = \frac{R_{mn}}{2\pi}
\end{equation}
where $R_{mn}$ is Ricci curvature of connections $w_{j}^{i}$. On
$\textbf{g}$ it is written as
\begin{equation} \label{equation3.147}
\beta_{ij} = \frac{1}{4\pi}f_{nj}^{k}f_{ik}^{n}
\end{equation}
On decomposed spaces $\textbf{h}$ and $\textbf{m}$ one loop
$\beta$-functions will be
\begin{align}
\beta_{ab} &= \frac{1}{4\pi}(f_{\beta b}^{\alpha}f_{a\alpha}^{\beta}
+ f_{db}^{c}f_{ac}^{d}) \label{equation3.148}\\
\beta_{\alpha\gamma} &=
\frac{1}{4\pi}(f_{\lambda\gamma}^{a}f_{\alpha a}^{\lambda} +
f_{\alpha\lambda}^{a}f_{a\gamma}^{\lambda}) \label{equation3.149}
\end{align}
It is readily observed that $R_{a\alpha} = R_{\alpha a} = 0$. On
pseudodual spaces one can write the following relations
\begin{equation} \label{equation3.150}
\beta_{ij} = - \tilde{\beta}_{ij} \ \ \ \ \ \ \ \ \ \beta_{ab} = -
\tilde{\beta}_{ab} \ \ \ \ \ \ \ \ \ \ \beta_{\alpha\gamma} = -
\tilde{\beta}_{\alpha\gamma}
\end{equation}
if there is a non-mixing pseudoduality. On the other hand if there
is a mixing pseudoduality we have
\begin{align}
\beta_{a b} = - \tilde{\beta}_{a b} - \tilde{\beta}_{a \nu}
T_{b}^{\nu}(0) - \tilde{\beta}_{\nu b} T_{a}^{v} (0) -
\tilde{\beta}_{\mu \nu} T_{a}^{\mu} (0) T_{b}^{\nu} (0)
\label{equation3.151}\\
\beta_{\mu \nu} = - \tilde{\beta}_{\mu \nu} - \tilde{\beta}_{d \nu}
T_{\mu}^{d}(0) - \tilde{\beta}_{\mu d} T_{\nu}^{d} (0) -
\tilde{\beta}_{a b} T_{\mu}^{a} (0) T_{\nu}^{b} (0)
\label{equation3.152}
\end{align}
where we defined $\tilde{\beta}_{\nu b} \equiv \frac{1}{2 \pi}
\{\tilde{R}_{\nu \mu b}^{c} T_{c}^{\mu}(0) + \tilde{R}_{\nu c
b}^{\mu} T_{\mu}^{c} (0)\}$, $\tilde{\beta}_{a \nu} \equiv
\frac{1}{2 \pi} \{\tilde{R}_{a \mu \nu}^{c} T_{c}^{\mu}(0) +
\tilde{R}_{a c \nu}^{\mu} T_{\mu}^{c} (0)\}$, $\tilde{\beta}_{d \nu}
\equiv \frac{1}{2 \pi} \{\tilde{R}_{d \lambda \nu}^{c}
T_{c}^{\lambda}(0) + \tilde{R}_{d c \nu}^{\lambda} T_{\lambda}^{c}
(0)\}$ and $\tilde{\beta}_{\mu d} \equiv \frac{1}{2 \pi}
\{\tilde{R}_{\mu \lambda d}^{c} T_{c}^{\lambda}(0) + \tilde{R}_{\mu
c d}^{\lambda} T_{\lambda}^{c} (0)\}$ on the contrary to
(\ref{equation3.146}). We notice that if all mixing isometries
vanish, then we get (\ref{equation3.150}). We notice that we will
also obtain additional mixing components of $\beta$-function, but we
avoid to obtain them.

\section{Discussion} \label{sec3:discussion}

In this section we were able to obtain infinite number of
pseudoduality equations by switching from Lie group expressions to
Lie algebra ones. We observed that pseudoduality transformation
respects the conservation law of currents. To understand what these
currents imply for let us write pseudoduality equations as
\begin{align}
\tilde{J}_{+}^{(L)} = +T J_{+}^{(L)} \notag\\
\tilde{J}_{-}^{(L)} = -T J_{-}^{(L)} \notag
\end{align}
where $J_{\pm}^{(L)} = g^{-1} \partial_{\pm}g$. First equation
implies that $T$ is a function of $\sigma^{+}$ as above. Second
equation is interesting and gives the information about currents. If
we take $\partial_{+}$ of second equation we obtain that
\begin{equation}
[\tilde{g}^{-1} \partial_{-} \tilde{g}, \tilde{g}^{-1} \partial_{-}
\tilde{g}]_{\tilde{G}} = - (\partial_{+} T) (g^{-1} \partial_{-} g)
- T [g^{-1} \partial_{-}g, g^{-1} \partial_{+}g]_{G} \notag
\end{equation}
We notice that $g^{-1} \partial_{\pm}g \in \textbf{g}$, and if we
use the definition $ad_{\textbf{g}} (X)(Y) = [X, Y]_{G}$ this
equation can be written as
\begin{equation}
ad_{\tilde{\textbf{g}}} (\tilde{J}_{+}^{(L)}) (\tilde{J}_{-}^{(L)})
= (\partial_{+} T) J_{-}^{(L)} + T ad_{\textbf{g}} (J_{+}^{(L)})
(J_{-}^{(L)}) \notag
\end{equation}
If the second pseudoduality equation is inserted then one gets
\begin{equation}
- ad_{\tilde{\textbf{g}}} (\tilde{J}_{+}^{(L)})T - T ad_{\textbf{g}}
(J_{+}^{(L)}) = (\partial_{+} T)  \notag
\end{equation}
It is obvious that this is the lie algebra version of the $AdG
\times Ad\tilde{G}$ action on T. $ad_{\textbf{g}}(J_{+}^{(L)})$ is
the orthogonal flat connection on $g^{*}TG$ as defined in section
(\ref{sec3:cur}). One may find curvature relations using these
connections as above. Thus another interpretation of pseudoduality
is that since $J_{+}^{(L)}$ depends only on $\sigma^{+}$, so does
$T$. Hence if we define a parallel transport $P(\sigma)$ from (0, 0)
to $\sigma = (\sigma^{+}, \sigma^{-})$, pseudoduality equations may
be written as
\begin{equation}
*_{\Sigma} (\tilde{P}(\sigma))^{-1} (\tilde{g}^{-1} d\tilde{g}) =
T(0) (P(\sigma)^{-1} g^{-1} dg) \notag
\end{equation}
where $T(0) = \tilde{P}(\sigma) T(\sigma) P^{-1} (\sigma)$. This
means that we start with $g^{-1} dg$, and parallel transport it to
origin, and do the same on the dual model. We finally use the fixed
isometry $T(0)$ to equate these two fields at the origins.

\section*{Acknowledgments}

I would like to thank O. Alvarez for his comments, helpful
discussions, and reading an earlier draft of the manuscript. I would
like to thank E. A. Ivanov for bringing his important paper to my
attention.

\bibliographystyle{amsplain}

\end{document}